\documentclass[prc,twocolumn,showpacs,twoside,showpacs,superscriptaddress]{revtex4-1}
\usepackage[colorlinks=true,linkcolor=blue]{hyperref}
\usepackage{amsmath,bbold,amssymb,epsfig,feynmp,color,ifpdf}
\usepackage{graphicx}
\usepackage{xcolor}
\usepackage{ulem}

\definecolor{dark-red}{rgb}{0.,0.,0}
\definecolor{dark-blue}{rgb}{0.,0.,1}
\definecolor{medium-blue}{rgb}{0,0,1}
\definecolor{gray}{rgb}{0.85,0.85,0.85}

\hypersetup{
    colorlinks, linkcolor={dark-red},
    citecolor={dark-blue}, urlcolor={medium-blue}
}

\begin{document}

\title{Finite-temperature linear response theory based on relativistic Hartree Bogoliubov model with point-coupling interaction}%

\author{A. Ravli\'c }%
\affiliation{Department of Physics, Faculty of Science, University of Zagreb, Bijeni\v{c}ka c. 32, 10000 Zagreb, Croatia}%
\author{Y.F. Niu}
\email{niuyf@lzu.edu.cn}
\affiliation{School of Nuclear Science and Technology, Lanzhou University, Lanzhou, China}
\author{T. Nik\v{s}i\'c}
\affiliation{Department of Physics, Faculty of Science, University of Zagreb, Bijeni\v{c}ka c. 32, 10000 Zagreb, Croatia}%
\author{N. Paar}
\affiliation{Department of Physics, Faculty of Science, University of Zagreb, Bijeni\v{c}ka c. 32, 10000 Zagreb, Croatia}%
\author{P. Ring}
\affiliation{Physik Department, Technische Universitat Munchen, D-85747 Garching, Germany}%

\date{\today}%
\revised{November 2020}%

\begin{abstract}
The finite-temperature linear response  theory based on the finite-temperature relativistic Hartree-Bogoliubov (FT-RHB) model is developed in the charge-exchange channel to study the temperature evolution of spin-isospin excitations. Calculations are performed self-consistently with relativistic point-coupling interactions DD-PC1 and DD-PCX. In the charge-exchange channel, the pairing interaction can be split into isovector ($T = 1$) and isoscalar ($T = 0$) parts. For the isovector component, the same separable form of the Gogny D1S pairing interaction is used both for the ground-state calculation as well as for the residual interaction, while the strength of the isoscalar pairing in the residual interaction is determined by comparison with experimental data on Gamow-Teller resonance (GTR) and  Isobaric analog resonance (IAR) centroid energy differences in even-even tin isotopes.  The temperature effects are introduced by treating Bogoliubov quasiparticles within a grand-canonical ensemble. Thus, unlike the conventional formulation of the quasiparticle random-phase approximation (QRPA) based on the Bardeen-Cooper-Schrieffer (BCS) basis, our model is formulated within the Hartree-Fock-Bogoliubov (HFB) quasiparticle basis. Implementing a relativistic point-coupling interaction and a separable pairing force allows for the reduction of complicated two-body residual interaction matrix elements, which considerably decreases the dimension of the problem in the coordinate space. The main advantage of this method is to avoid the diagonalization of a large QRPA matrix, especially at finite temperature where the size of configuration space is significantly increased. The implementation of the linear response code is used to study the temperature evolution of IAR, GTR, and spin-dipole resonance (SDR) in even-even tin isotopes in the temperature range $T = 0 - 1.5$ MeV.
\end{abstract}

\maketitle

\section{Introduction}

Spin-isospin excitations play an important role not only in understanding isovector terms of the effective nucleon-nucleon interaction  and the symmetry energy in nuclear matter~\cite{Paar2007_RPP70-691},
but also in calculating reaction rates mediated by the weak interaction such as the charged lepton capture, $\beta-$decay rates and neutrino-nucleus reactions and scattering,  which are of significance for understanding the nucleosynthesis of elements heavier than iron in the r-process~\cite{Kajino2019_PPNP107-109} and the evolution of core-collapse supernovae~\cite{Janka2012_ARNPS52-407,Janka2007_PR442-38}.

The spin-isospin excitations of the charge-exchange type occur between neighbouring nuclei in either $\beta^+$ or $\beta^-$ direction, including the isobaric analogue state (IAS), Gamow-Teller (GT) transitions, and spin-dipole (SD) transitions~\cite{Osterfeld1992_RMP64-491}. Experimentally, the spin-isospin excitations can be studied with the charge-exchange reactions such as $(p,n)$ or $({}^3\text{He},t)$ in $\beta^-$ direction~\cite{Ichimura2006_PPNP56-446}, or spontaneously via $\beta-$decay~\cite{Hatori1992_NPA549-327}.

Theoretically, the spin-isospin response of the charge-exchange type can be investigated by the shell-model (SM) approach  and proton-neutron random-phase approximation (RPA). Although SM calculations can provide an excellent agreement with the experimental data~\cite{WANG-LJ2018_PRC97-044302,Suzuki2003_PRC67-044302,Suzuki2011_PRC83-044619,Caurier2005_RMP77-427}, at present, they are limited to the mass range of $A \leq 70$ or nuclei around the magic numbers due to the huge configurational space involved. The RPA approach represents a linearization of the time-dependent Hartree-Fock (TDHF) equation~\cite{Ring1980} and provides a unified description of nuclei from the valley of $\beta-$stability all the way to the nucleon drip-lines employing only a small number of parameters adjusted to basic nuclear properties~\cite{Niksic2008_PRC78-034318,RocaMaza2018_PPNP101-96,Lalazissis2005_PRC71-024312,Paar2004_PRC69-054303}. Based on relativistic~\cite{Paar2004_PRC69-054303,Litvinova2014_PLB730-307,NIU-ZM2017_PRC95-044301,
PenaArteaga2008_PRC77-034317,Daoutidis2009_PRC80-024309} and non-relativistic~\cite{RocaMaza2013_PST154-014011,Bender2002_PRC65-54322,Fracasso2007_PRC76-044307,NIU-YF2012_PRC85-034314,
Mustonen2016_PRC93-014304} energy density functionals (EDFs), self-consistent RPA and quasiparticle RPA (QRPA) that includes pairing correlations have been realized, with an excellent extrapolation ability.

% something about finite-temperature spin-isospin
The temperature evolution of spin-isospin excitations is of particular relevance for nuclear astrophysics, especially for the initial stages of the r-process where the temperature evolves in the range $10^9 - 10^{10}$ K~\cite{Mumpower2016_PPNP86-86}, and in the final stages of the core-collapse supernovae where the temperature can reach up to $3 \times 10^{10}$ K~\cite{Sullivan2015_ApJ816-44}. To describe this, temperature effects have to be considered in the RPA or QRPA approach and self-consistently in the ground state calculations like the Hartree-Fock (HF) or HF + Bardeen-Cooper-Schrieffer (BCS), or Hartree-Fock-Bogoliubov (HFB) approaches. A fully self-consistent framework for the calculation of electron-capture rates was developed in Refs.~\cite{Paar2009_PRC80-055801,Fantina2012_PRC86-035805} based on the finite-temperature proton-neutron RPA (FT-PNRPA) using Skyrme EDFs. Based on relativistic density functionals, the FT-PNRPA approach was developed in Ref.~\cite{NIU-YF2011_PRC83-045807} and also applied to the calculation of electron-capture rates. The above FT-PNRPA approaches did not include pairing correlations in open-shell nuclei which are important at temperatures below the pairing collapse. The finite-temperature quasiparticle RPA (FT-QRPA) based on the finite-temperature Hartree-Fock + BCS theory using Skyrme density functionals was developed in Refs.~\cite{Yuksel2017_PRC96-024303,Yuksel2019_EPJA55-1} and applied to non-charge-exchange multipole excitations and in Ref.~\cite{Yuksel2020_PRC101-044305} to the calculation of the temperature evolution of the GT${}^-$ strength within the proton-neutron FT-QRPA (FT-PNQRPA) formalism. Within relativistic functionals, FT-PNQRPA based on the FT-Hartree+BCS (FT-HBCS) model was used for the calculation of electron-capture rates and the temperature evolution of $\beta-$decay half-lives~\cite{Ravlic2020_PRC102-065804,Ravlic2020_arXiv2010.06394}.

Covariant density functional theory (CDFT)  is based on Lorentz invariance, which connects in a consistent way the spin and spatial degrees of freedom of the nucleus. It has achieved great success in describing the nuclear ground-state and excited-state properties with a small number of parameters~\cite{Boguta1977_NPA292-413,Ring1996_PPNP37-193,Meng2006_PPNP57-470}. Within CDFT, there are two representations of the effective nuclear interaction: one is the finite-range meson-exchange representation, where the nucleons interact with each other through the exchange of mesons, such as the isoscalar scalar meson $\sigma$, the isoscalar vector meson $\omega$, and the isovector vector meson $\rho$ as well as the electromagnetic field. The other approach is the point coupling representation, where local contact interactions between the nucleons replace the meson exchange in each channel. This is justified by the large masses of the mesons and the corresponding short range of the forces~\cite{Lalazissis2005_PRC71-024312,Niksic2008_PRC78-034318}. For both effective interactions, the medium dependence is necessary for a quantitative treatment of nuclear matter and finite nuclei, which can be considered  by including nonlinear terms~\cite{Boguta1977_NPA292-413} or by assuming an explicit density dependence for the coupling constants~\cite{Lalazissis2005_PRC71-024312,Niksic2008_PRC78-034318}.
The point-coupling effective interaction becomes more and more popular in complicated calculations due to its simplicity and at the same time good performance as compared to the meson-exchange effective interaction. For example, the interaction DD-PC1~\cite{Niksic2008_PRC78-034318} and the newly developed DD-PCX~\cite{Yuksel2019_PRC99-034318} have shown their success in describing nuclear ground-state and excited-state properties~\cite{Vale2021_PRC103-064307}.

Previously introduced self-consistent FT-QRPA calculations are all based on finite-temperature BCS approaches. However, the BCS model faces serious problems in nuclei with a large neutron excess in the neighborhood of the drip line where the Fermi level is close to the continuum. In these cases, one has to use the Hartree-Fock-Bogoliubov theory where pairing correlations in the continuum are treated in a more consistent way~\cite{Dobaczewski1984_NPA422-103}. Within the mean-field approximation, the essential effect of introducing finite temperature is to produce a Fermi-Dirac distribution for these independent particles. In the framework of finite-temperature HFB (FT-HFB), the independent particles are Bogoliubov quasiparticles that obey the Fermi-Dirac distribution at a certain temperature. Therefore, starting from the FT-HFB, it is straight-forward to build the QRPA equation directly in the Bogoliubov quasiparticle basis~\cite{Sommermann1983_APNY151-163,Ring1984_NPA419-261}. In the realistic case it leads to the complexity of dealing with the full HFB-wave functions $U$ and $V$, and a diagonalization problem of considerable dimension.

Usually, for zero and for finite temperatures, the QRPA equation  is written in the HF-BCS basis which largely simplifies the calculations, because in this case the two-body matrix elements $V_{1234}$ are calculated in the particle basis of the HF-solution and the pairing properties enter only through specific combinations of the BCS-occupation amplitudes. In the QRPA based on the HFB solution (HFB-QRPA) the QRPA matrix is written in Bogoliubov quasiparticle space with the matrix elements $H^{22}_{1234}$ and $H^{40}_{1234}$~\cite{Ring1980}. These are not only as twice as many matrix elements, but they also involve the full HFB-wave functions $U$ and $V$. By this reason, in the literature, the full HFB-QRPA equations in Bogoliubov quasiparticle space (with or without temperature), have rarely been solved for realistic density functionals~\cite{Giambrone2003_NPA726-3,Khan2004_NPA731-311}.

For vanishing temperature the full HFB-QRPA equations can be solved in the canonical basis~\cite{Paar2003_PRC67-034312,Terasaki2005_PRC71-034310}, because in this basis the HFB wavefunction for the ground state has BCS form and this method is used in all the present HFB-QRPA applications. However this method cannot be extended so easily for finite temperatures, because in the canonical basis the Hamiltonian is not diagonal and therefore the corresponding quasiparticles do not form an independent statistical ensemble. Neglecting this fact and using, in the corresponding QRPA equations, Boltzmann factors containing the BCS-expression in the canonical basis, is only an approximation. Its validity has to be investigated.

As in all RPA or QRPA calculations, the corresponding eigenmodes can be determined either by diagonalizing the RPA/QRPA matrix in an appropriate basis or by solving the linear response equations in a time-dependent external field~\cite{Ring1980}. In the first case, one obtains in one diagonalization all the eigenmodes and the corresponding wave functions of the system, whereas in the second case one has to solve the linear response equation by inverting this matrix for a mesh in the energy space and it is relatively easy to derive the response function for various external fields. The wavefunctions and energies of specific discrete states are obtained by contour integrations in the complex energy plane. These two methods lead to exactly identical results. However, there are cases where one of them is clearly preferable. The linear response method is definitely preferable
in the case of continuum RPA for zero range forces \cite{Bertsch1973_PRL31-121,Daoutidis2009_PRC80-024309} or for applications beyond mean field, where the effective integral kernel depends on the energy~\cite{Litvinova2007_PRC75-064308}.

In this investigation we present a relatively simple and precise method for the full solution of the temperature dependent HFB-QRPA equations for spherical nuclei with realistic density functionals of zero range. We apply it for proton-neutron QRPA with covariant functionals, but it is definitely also applicable for other relativistic and non-relativistic cases. For this goal we use (i) the linear response formalism with all the advantages discussed above and (ii) we represent the zero range force as a sum of separable terms. This not only allows to reduce considerably the dimension of the matrix to be inverted, but it also simplifies essentially the transformation of the matrix elements to quasiparticle space, because here we have to deal with a finite number of one-body instead of two-body operators. The basic ideas of this method have been presented already in Ref.~\cite{Ring1984_NPA419-261} for a schematic model for the description of giant resonances in deformed rotating nuclei at finite temperature. Even in this relatively simple model the solution of the temperature dependent QRPA equations was by no means trivial, because all the symmetries are broken in this case, which leads to matrices of very large dimensions.

The advantage of separable forces can be also used in the particle-particle channel. In this case, zero-range forces, as they are commonly used in non-relativistic Skyrme calculations~\cite{Bender2000_EPJA8-59} have severe problems because of their ultra-violet divergence~\cite{Karatzikos2010_PLB689-72}. We, therefore, use a separable version of the finite range Gogny force D1S~\cite{Berger1991_CPC63-365} proposed by Tian et al. for the ground state~\cite{Tian2009_PLB676-44} and for the QRPA calculations~\cite{TIAN-Y2009_PRC79-064301}.

In this work, we develop a self-consistent FT-PNQRPA in the charge-exchange channel, based on the finite-temperature relativistic Hartree-Bogoliubov (FT-RHB) model using point-coupling EDFs: DD-PC1~\cite{Niksic2008_PRC78-034318} and DD-PCX~\cite{Yuksel2019_PRC99-034318}, within the linear response formalism.  The pairing correlations are treated  with the separable pairing interaction \cite{Tian2009_PLB676-44,TIAN-Y2009_PRC79-064301} both for the ground-state and for the QRPA. For the calculation of the nuclear ground state, the finite-temperature relativistic Hartree-Bogoliubov theory (FT-RHB) developed in Ref.~\cite{NIU-YF2013_PRC88-034308} is used. We note that in the calculation of the ground-state, no proton-neutron mixing is assumed. Hence only isovector ($T=1$) pairing contributes, while for spin-isospin excitations, the type of included pairing interaction depends on the parity of the transition. For considered natural parity transitions ($J^\pi = 0^+, 1^-$), same isovector ($T=1$) pairing is employed as in ground-state, while for unnatural parity transitions ($J^\pi = 0^-, 1^+, 2^-$), isoscalar pairing ($T = 0$) interaction of the same form is used. At present, there is no consensus about the strength of the isoscalar pairing interaction. However, some studies imply that it should be of the same magnitude as the isovector pairing~\cite{BAI-CL2014_PRC90-054335,Sagawa2016_PS91-083011}.

The paper is organized as follows. In Sec. \ref{sec:theoretical_formalism} we present a brief introduction to the FT-RHB theory and linear response FT-PNQRPA supplemented with additional derivations in the appendices \ref{sec:app_a} , \ref{sec:appb} and \ref{sec:appc}. Details on the numerical implementation and tests with available codes in matrix formulation can be found in Sec. \ref{sec:numerical_implementations_and_tests}. Illustrative calculations of the temperature evolution of spin-isospin excitations for even-even tin isotopes are presented in Sec. \ref{sec:illustrative_calculations}. Finally, Sec. \ref{sec:conclusion} contains concluding remarks and an outlook.

\section{Theoretical formalism}\label{sec:theoretical_formalism}

\subsection{Finite-temperature Hartree Bogoliubov theory}

At finite-temperature the nucleus is treated within the grand-canonical ensemble, being in equilibrium with a heat bath of temperature $T$ with chemical potential $\lambda$. The ground-state eigenvalue problem can be obtained by the variational principle from the grand-canonical potential~\cite{Goodman1981_NPA352-30}
\begin{equation}\label{Omega}
\Omega = E - TS - \lambda N,
\end{equation}
where $E$ is the ground-state energy, $S$ entropy and $N$ particle number. For superfluid systems the variation is performed in the space of Bogoliubov quasiparticles
\begin{align}
\beta^\dag_k &= \sum^M_{l=1} U_{lk}c^\dag_l+V_{lk}c^{}_l.\\
\beta^{}_k &= \sum^M_{l=1} V^\ast_{lk}c^\dag_l+ U^\ast_{lk}c_l.
\end{align}
It is convenient to introduce the set of operators $a_\mu$ which combine creation and annihilation quasiparticle (q.p.) operators $\beta_k^\dag$ and $\beta_k$ as in Ref.~\cite{Ring1984_NPA419-261}
\begin{equation}\label{eq:definition_of_operators}
\left.
\begin{array}{l}a_{\mu}=\beta_{k} \\ a_{\bar{\mu}}=\beta_{k}^{\dag}
\end{array}\right\} \quad k=1 \ldots M ; \mu=1 \ldots M,-1,-2, \ldots,-M,
\end{equation}
and obey the commutation relations
\begin{equation}
\{ a_\mu, a_{\mu^\prime} \} = \delta_{\mu \bar{\mu}^\prime},
\end{equation}

The corresponding generalized density matrix is the Valatin-density~\cite{Valatin1961_PR122-1012} of dimension $2M\times 2M$:
\begin{equation}\label{Valatin-density}
{\mathcal{R}} =  \begin{pmatrix}
\langle \beta^\dag_{k^\prime} \beta_k \rangle_T & \langle \beta_{k^\prime} \beta_k \rangle_T \\
\langle \beta^\dag_{k^\prime} \beta^\dag_k \rangle_T & \langle \beta_{k^\prime} \beta^\dag_k \rangle_T \\
\end{pmatrix} =
\begin{pmatrix}
\rho_{kk^{\prime}} & \kappa_{kk^{\prime}} \\
-\kappa^\ast_{kk^{\prime}} & 1-\rho^\ast_{kk^{\prime}}\\
\end{pmatrix},
\end{equation}
where, at finite temperature $T$, $\langle \cdot \rangle_T$ denotes the thermal average. At finite temperature, in the statistical ensemble of independent quasiparticles this matrix has the form of a Fermi-Dirac distribution
\begin{equation}
{\mathcal{\hat R}} = Z^{-1} e^{-\beta ({\mathcal{\hat H}} - \lambda {\mathcal{\hat N}})},
\end{equation}
where $\mathcal{H}_{\mu \mu^\prime} = \langle \{ [a_\mu, H], a_{\mu^\prime}^\dag \} \rangle_T$ is the mean-field hamiltonian, $Z = \text{Tr} [e^{-\beta ({\mathcal{\hat H}} - \lambda {\mathcal{\hat N}})}]$ is the grand-canonical partition function, and $\beta = 1/k_B T$ with $k_B$ being the Boltzmann constant.

The ground state is obtained by the variation of the grand-canonical potential (\ref{Omega})
in quasiparticle space with respect to the density ${\mathcal{R}}$:
\begin{equation}
\frac{\delta \Omega}{\delta {\mathcal{R}}} = 0.
\end{equation}
This leads to the finite-temperature relativistic Hartree-Bogoliubov (FT-RHB) equations~\cite{Goodman1981_NPA352-30,NIU-YF2013_PRC88-034308}
\begin{equation}
\left(
\begin{array}{cc}
h-\lambda-M & \Delta \\
-\Delta^{*} & -h^{*}+\lambda+M
\end{array}
\right)
\left(\begin{array}{l}
U_k \\ V_k
\end{array}
\right)
= E_k \left(\begin{array}{l}
U_k \\ V_k
\end{array}
\right),
\end{equation}
where $h$ is the mean-field Dirac Hamiltonian and $\Delta$ is the pairing filed describing the particle-particle correlations, the nucleon mass is denoted by $M$, and the chemical potential $\lambda$ is determined by the particle number subsidiary condition $\langle\hat{N}\rangle = \text{Tr}[\rho] = N$,
where $N$ is either the proton or the neutron particle number.  $E_k$ denote the q.p. energies and $U_k,V_k$ are the corresponding RHB wavefunctions.
In this basis the generalized density is diagonal
\begin{equation}\label{Valatin-density}
{\mathcal{R}} =
\begin{pmatrix}
f_k & 0 \\
0 & 1-f_k \\
\end{pmatrix},
\end{equation}
and $f_k$ is the Fermi-Dirac factor
\begin{equation}
f_{k}=\frac{1}{1+e^{\beta E_{k}}}.
\end{equation}
In the relativistic case~\cite{Kucharek1991_ZPA339-23} the HFB wavefunctions $U_k$ and $V_k$  have the form of Dirac spinors.
The single-particle Dirac Hamiltonian $h$ is given by
\begin{equation}
h=\boldsymbol{\alpha} \cdot \boldsymbol{p}+V(\boldsymbol{r})+\beta(M+S(\boldsymbol{r})),
\end{equation}
where $\boldsymbol{p}$ is the nucleon momentum, $V$ is the time-like component of the vector field and $S$ is the scalar field. For the relativistic point-coupling interactions used here, they can be written as~\cite{Niksic2008_PRC78-034318}
\begin{equation}
S(\boldsymbol{r}) = \alpha_S \rho_s + \delta_S \nabla^2\rho_s,
\end{equation}
\begin{equation}
V(\boldsymbol{r})=\alpha_{V} \rho_{v}+\alpha_{T V} \tau_{3} \rho_{t v}+e A_{0}+\Sigma_{0}^{R},
\end{equation}
where $\rho_s, \rho_v$ and $\rho_{tv}$ are scalar, vector and isovector densities, respectively. $\alpha_S, \alpha_{V}, \alpha_{TV} $ are the density-dependent couplings depending on $\rho_v$, $A_0$ is the time component of the electromagnetic field and $\Sigma_0^R$ is the rearrangement contribution
\begin{equation}
\Sigma_0^R = \frac{\partial \alpha_S}{\partial \rho_v} \rho_s^2 + \frac{\partial \alpha_V}{\partial \rho_v} \rho_v^2 + \frac{\partial \alpha_{TV}}{\partial \rho_v} \rho_{tv}^2,
\end{equation}
with functional form of couplings being defined in Refs.~\cite{Niksic2008_PRC78-034318,Daoutidis2009_PRC80-024309}.
The scalar, vector and isovector densities are calculated within FT-RHB theory~\cite{Ring1984_NPA419-261} as
\begin{equation}
\begin{aligned} \rho_{s} &=\sum_{E_{k}>0} V_{k}^{\dagger} \gamma^{0}\left(1-f_{k}\right) V_{k}+U_{k}^{T} \gamma^{0} f_{k} U_{k}^{*} \\
 \rho_{v} &=\sum_{E_{k}>0} V_{k}^{\dagger}\left(1-f_{k}\right) V_{k}+U_{k}^{T} f_{k} U_{k}^{*} \\
 \rho_{tv } &=\sum_{E_{k}>0} V_{k}^{\dagger} \tau_{3}\left(1-f_{k}\right) V_{k}+U_{k}^{T} \tau_{3} f_{k} U_{k}^{*}  \end{aligned},
\end{equation}
where $\tau_3$ is the third component of the Pauli isospin matrix.

Within this work two parameter-sets of the relativistic point-coupling interactions will be employed: DD-PC1 \cite{Niksic2008_PRC78-034318} and DD-PCX~\cite{Yuksel2019_PRC99-034318}.

The pairing field is calculated as
\begin{equation}
\Delta_{l l^{\prime}}=\frac{1}{2} \sum_{k k^{\prime}} V_{l l^{\prime} k k^{\prime}}^{p p} \kappa_{k k^{\prime}},
\end{equation}
where $V^{pp}$ is the matrix element of the particle-particle ($pp$) interaction~\cite{Tian2009_PLB676-44} and $\kappa$ is the pairing tensor
\begin{equation}
\kappa= \sum \limits_{E_k > 0 } V_k^{*}(1-f_k) U_k^{T}+U_k f_k V_k^{\dagger}.
\end{equation}

The mean pairing gap $\Delta$ is then defined as
\begin{equation}
\Delta = \frac{\sum_{l l^\prime} \Delta_{l l^\prime} \kappa_{l l^\prime}}{\sum_l \kappa_{ll}}.
\end{equation}
For the $pp$ interaction $V^{pp}$ we adopt the separable interaction of the form\cite{Tian2009_PLB676-44,TIAN-Y2009_PRC79-064301}
\begin{equation}\label{eq-15}
V\left(\mathbf{r}_{1}, \mathbf{r}_{2}, \mathbf{r}_{1}^{\prime}, \mathbf{r}_{2}^{\prime}\right)=-G \delta\left(\mathbf{R}-\mathbf{R}^{\prime}\right) P(r) P\left(r^{\prime}\right) \frac{1}{2}\left(1-P^{\sigma}\right),
\end{equation}
where $\boldsymbol{R} = \frac{1}{2}(\boldsymbol{r}_1 + \boldsymbol{r}_2)$ and $\boldsymbol{r} = \boldsymbol{r}_1 - \boldsymbol{r}_2$ are the center-of-mass and relative coordinates respectively, and $P(r)$ is defined as
\begin{equation}\label{eq-16}
P(r)=\frac{1}{\left(4 \pi a^{2}\right)^{3 / 2}} e^{-\frac{r^{2}}{4 a^{2}}}.
\end{equation}
The parameters of the $pp$ interaction are $G_p = G_n = 728$ MeV fm${}^3$ for DD-PC1 interaction ~\cite{Niksic2008_PRC78-034318} and $G_p =773.78 $ MeV fm${}^3$ and $G_n = 800.66 $ MeV fm${}^3$ for DD-PCX~\cite{Yuksel2019_PRC99-034318} for protons and neutrons respectively, while $a^2=0.644$ fm${}^2$  for both interactions.
\subsection{Linear response theory at finite temperature}

The linear response equation is derived by introducing the time-dependent external field $\mathcal{F}(t)$ on top of the FT-RHB ground-state. In this case the density $\mathcal{R}(t)$ depends on time and it obeys the equation of motion
\begin{equation}\label{eq:tdhf}
i\dot{\mathcal{R}}(t) = [\mathcal{H}(\mathcal{R}(t))+\mathcal{F}(t), \mathcal{R}(t)].
\end{equation}
If there is no external field, the above equation reduces to the FT-RHB equation
\begin{equation}
[\mathcal{H}(\mathcal{R}^0), \mathcal{R}^0] = 0,
\end{equation}
with the static solution $\mathcal{R}^0$ at finite temperature. In this basis $\mathcal{R}^0$ and $\mathcal{H}(\mathcal{R}^0)$ are diagonal, with the eigenvalues
\begin{equation}
\left.\begin{array}{ll}f_{\mu}=f_{k}, & f_{\bar{\mu}}=1-f_{k} \\ E_{\mu}=E_{k}, & E_{\bar{\mu}}=-E_{k}\end{array}\right\} \quad \mu>0.
\end{equation}

By linearizing the generalized density
\begin{equation}
\mathcal{R}(t)=\mathcal{R}^{0}+\left(\delta \mathcal{R} e^{-i E t}+h . c .\right),
\end{equation}
and inserting in Eq. (\ref{eq:tdhf}) the linear response equation for charge-changing transitions is obtained
\begin{align}\label{eq:lin_res_matrix_xy}
\begin{split}
&\left(\omega-E_{\pi}+E_{\nu}\right) \delta \mathcal{R}_{\pi \nu} \\
&= \left(f_{\nu}-f_{\pi}\right)\left\{F_{\pi \nu}+ \sum_{\pi^\prime \nu^{\prime}} \mathbb{W}_{\pi \nu \pi^\prime \nu^{\prime}} \delta \mathcal{R}_{\pi^\prime \nu^{\prime}}\right\},
\end{split}
\end{align}
where ($\pi$) and   ($\nu$) are proton-quasiparticle and neutron-quasiparticle states.
The matrix elements of the external field operator $\mathcal{F}$ are denoted by $F_{\pi \nu}$ and the effective interaction $\mathbb{W}_{\pi \nu \pi^{\prime} \nu^{\prime}}$ is defined as
\begin{equation}\label{eq:effective_interacton_W}
\mathbb{W}_{\pi \nu \pi^\prime \nu^\prime} = \frac{\delta \mathcal{H}_{\pi \nu}}{\delta \mathcal{R}_{\pi^\prime \nu^\prime}}.
\end{equation}
Introducing the response function $\mathbb{R}$~\cite{Ring1984_NPA419-261,Ring1980}
\begin{equation}
\delta \mathcal{R}_{\pi \nu}= \sum_{\pi^{\prime} \nu^{\prime}} \mathbb{R}_{\pi \nu \pi^{\prime} \nu^{\prime}} F_{\pi^{\prime} \nu^{\prime}},
\end{equation}
we obtain the Bethe-Salpeter equation
\begin{equation}\label{eq:bethe_salpeter_qp}
\mathbb{R}_{\pi \nu \pi^{\prime} \nu^{\prime}}=\mathbb{R}_{\pi \nu \pi^{\prime} \nu^{\prime}}^{0}+ \sum_{\mu \mu^{\prime}} \mathbb{R}_{\pi \nu \mu \mu^{\prime}}^{0} \sum_{\pi^{\prime \prime} \nu^{\prime \prime}} \mathbb{W}_{\pi \nu \pi^{\prime \prime} \nu^{\prime \prime}} \mathbb{R}_{\pi^{\prime \prime} \nu^{\prime \prime} \pi^{\prime} \nu^{\prime}}.
\end{equation}
By setting the interaction term to zero we obtain the unperturbed response
\begin{equation}
\mathbb{R}_{\pi \nu \pi^{\prime} \nu^{\prime}}^{0}=\frac{\left(f_{\pi}-f_{\nu}\right)}{\omega-E_{\pi}-E_{\nu}+i \eta} \delta_{\pi \pi^{\prime}} \delta_{\nu \nu^{\prime}},
\end{equation}
where the small parameter $i \eta$ has been added due to the analytic structure of the response function~\cite{Ring1984_NPA419-261,Ring1980}. As discussed in the introduction, for point-coupling interactions, the full Hamiltonian $\hat{H}$ can be written as a separable form~\cite{Daoutidis2009_PRC80-024309}
\begin{equation}\label{eq:full_ham}
\hat{H}=\hat{H}_{0}+ \sum_{\rho} \chi_\rho D_{\rho}^{\dagger} D_{\rho},
\end{equation}
where $\hat{H}_0$ is the mean-field Hamiltonian, $\rho$ runs over a set of single-particle operators $D_\rho$, and $\chi_\rho$ represents the coupling of the residual interaction channel. As discussed in Ref~\cite{Daoutidis2009_PRC80-024309}, $\rho$ runs for the point-coupling models over the various relativistic channels and over the mesh points in $r$-space. For this separable form of the interaction~\cite{Ring1984_NPA419-261} we keep, as usual in RMF-models, only the direct terms
\begin{equation}\label{eq:W_matrix}
\mathbb{W}_{\pi \nu \pi^{\prime} \nu^{\prime}}= \sum \limits_\rho \chi_\rho \mathcal{D}^*_{\rho_{\pi \nu}} \mathcal{D}_{\rho_{\pi^{\prime} \nu^{\prime}}} + \chi_\rho \mathcal{D}^*_{\rho_{\bar{\pi}^{\prime} \bar{\nu}^{\prime}}} \mathcal{D}_{\rho_{\bar{\pi} \bar{\nu}}}.
\end{equation}
This considerably simplifies the linear response equation. The above two terms can be effectively treated as two separate channels. Instead of solving the Bethe-Salpeter equation in quasiparticle space, as in Eq. (\ref{eq:bethe_salpeter_qp}), we introduce the reduced response function as
\begin{equation}
R_{\rho \rho^\prime}(\omega) = \sum \limits_{\pi \nu \pi^\prime \nu^\prime} \mathcal{D}^*_{\rho_{\pi \nu}}\mathbb{R}_{\pi \nu \pi^\prime \nu^\prime}(\omega) \mathcal{D}_{\rho^{\prime}_{\pi^\prime \nu^\prime}}.
\end{equation}
The unperturbed reduced response function is given by a simple substitution of $\mathbb{R}^0$ in the above definition
\begin{equation}\label{eq:reduced_unp_resp}
R_{\rho \rho^{\prime}}^{0}= \sum_{\pi \nu} \frac{\mathcal{D}_{\rho_{\pi \nu}}^{*} \mathcal{D}_{\rho_{\pi \nu}^{\prime}}\left(f_{\nu}-f_{\pi}\right)}{\omega-E_{\pi}+E_{\nu}+i \eta},
\end{equation}
which now yields the reduced Bethe-Salpeter equation in the $\rho$-space (i.e. in $r$-space)
\begin{equation}\label{eq:bethe_salpeter_coord_space}
\begin{aligned} R_{\rho \rho^{\prime}} &=R_{\rho \rho^{\prime}}^{0}+\sum_{\rho^{\prime \prime}} R_{\rho \rho^{\prime \prime}}^{0} \chi_{\rho^{\prime \prime}} R_{\rho^{\prime \prime} \rho^{\prime}} \end{aligned}.
\end{equation}
This equation presents a linear equation for the unknown matrices $R_{\rho \rho^{\prime}}$ in $\rho$-space. It is solved by inversion of the matrix
$\delta_{\rho\rho^{\prime}}-R_{\rho \rho^{\prime}}^{0} \chi_{\rho^{\prime}}$ whose dimension is the number of separable terms in the expansion (\ref{eq:full_ham}).
The strength function can be calculated as~\cite{Ring1984_NPA419-261,Ring1980}
\begin{equation}\label{eq:strength_function}
\begin{aligned} S_{F}(\omega) &=-\frac{1}{\pi}  \operatorname{Im}\left(\sum_{\pi \nu \pi^{\prime} \nu^{\prime}} F_{\pi \nu}^{*} \mathbb{R}_{\pi \nu \pi^\prime \nu^{\prime}}(\omega) F_{\pi^\prime \nu^{\prime}}\right) \end{aligned}.
\end{equation}

In the following we will denote the separable terms of particle-hole residual interaction from relativistic point-coupling density functinal as $Q_{c \pi \nu}(r)$ where $c$ is the channel index. If the external field can be written in terms of separable channels $Q_{c \pi \nu}(r)$, then~\cite{Daoutidis2009_PRC80-024309,Daoutidis2011_PRC83-044303}
\begin{equation}
F_{\pi \nu}=\sum_{c} \int r^{2} d r f_{c}(r) Q_{c \pi \nu}(r),
\end{equation}
where the radial dependence of $F_{\pi \nu}$ is contained in $f_c(r)$. We note that this is the case for the spin-isospin excitations considered within this work. From Eq. (\ref{eq:strength_function}) we obtain the strength function as
\begin{align}
\begin{split}
S_F(\omega) &= - \frac{1}{\pi} \text{Im} \left(\sum_{c c^{\prime}} \int r^{2} d r \int r^{\prime 2} d r^{\prime} f_{c}^{*}(r) R_{c c^{\prime}}\left(r, r^{\prime}\right) f_{c^{\prime}}\left(r^{\prime}\right) \right).
\end{split}
\end{align}
If $F_{\pi \nu}$ cannot be expressed in terms of the separable interaction channels $Q_{c \pi \nu}(r)$, then an additional step in solving the Bethe-Salpeter equation is needed as described in Ref.~\cite{Daoutidis2009_PRC80-024309}. If we define the response function for the external field operator $\hat{F}$ as
\begin{equation}
R_{FF} = \sum \limits_{\pi \nu \pi^\prime  \nu^\prime} F^*_{\pi \nu} \mathbb{R}_{\pi \nu \pi^\prime \nu^\prime} F_{\pi^\prime \nu^\prime},
\end{equation}
then the discrete FT-QRPA strength can be calculated using the contour integral~\cite{Daoutidis2009_PRC80-024309,Hinohara2013_PRC87-064309}
\begin{equation}\label{eq:contour_strength}
B(\hat{F}) \equiv |\langle i | \hat{F} | 0 \rangle|^2 = \frac{1}{2 \pi i} \oint \limits_{C_i} R_{FF} (\omega) d \omega,
\end{equation}
for the FT-QRPA eigenvalue $i$, where the $C_i$ is an appropriately chosen contour in the complex energy plane that encloses the $i$-th pole. Details regarding the calculation of discrete FT-QRPA modes within the linear response theory are given in Appendix \ref{sec:appc}. Having the FT-QRPA modes $P^i_{\pi \nu}, X^i_{\pi \nu}, Y^i_{\pi \nu}, Q^i_{\pi \nu}$ the contribution of particular 2 q.p. excitation in $\beta^-$ direction is obtained as
\begin{align}\label{eq:ftqrpa_excitation_me}
\begin{split}
\langle i | \hat{F} | 0 \rangle_{\pi \nu} &= -P^i_{\pi \nu} (U^\dag F U)_{\pi \nu} + X^i_{\pi \nu} (U^\dag F V^*)_{\pi \nu} \\
&- Y^i_{\pi \nu} (V^T F U)_{\pi \nu} + Q^i_{\pi \nu} (V^T F V^*)_{\pi \nu},
\end{split}
\end{align}
with respect to the external field operator $\hat{F}$.

\subsection{Separable channel matrix elements in proton-neutron quasiparticle basis}

For the residual particle-hole ($ph$) interaction the separable terms, coupled to a good angular momentum $J$ and projection $M$ can be written in the proton-neutron quasiparticle basis as
\begin{align}
\begin{split}
\hat{Q}_{pn} &= \sum \limits_{\pi \nu} \begin{pmatrix}
U^\dag Q^J U & U^\dag Q^J V^* \\
V^T Q^J U & -V^T Q^J V^*
\end{pmatrix}_{\pi \nu} [a_\pi^\dag a_\nu]_{J M},  \\
\end{split}
\end{align}
where the operators $a^\dag$, $a$  are defined in Eq. (\ref{eq:definition_of_operators}). %Labels $S$ and $L$ are total spin and orbital angular momentum of the excitation respectively.
The $ph$ matrix elements $Q^{J}$ for the point-coupling interactions employed can be found in Appendix \ref{sec:app_a}.
Similarly, for the particle-particle residual interaction ($pp$) it follows
\begin{align}\label{eq:pp_pn}
\begin{split}
\hat{V}_{pn} &= \sum \limits_{\pi \nu} \begin{pmatrix}
-U^\dag V^J V & U^\dag V^J U^* \\
-V^T V^J V & -V^T V^J U^*
\end{pmatrix}_{\pi \nu} [a_\pi^\dag a_\nu]_{J M}, \\
\end{split}
\end{align}
where the matrix elements of separable $pp$ interaction are given in Eq. (\ref{eq:separable_pairing}). More details about the derivation and angular momentum coupling, with the definition of $[a_\pi^\dag a_\nu]_{J M}$ can be found in Appendix \ref{sec:appb}. Here we note that the second term in Eq. (\ref{eq:W_matrix}) is treated as another separable channel, thus the total number of channels $N_\rho$ has to be multiplied by 2 to account for all the terms in proton-neutron basis.

The external field operator for the excitation strength in the $\beta^-$ direction, assuming spherical symmetry, can be written as
\begin{align}\label{eq:external_matrix_element}
\begin{split}
\hat{F}_{pn} &= \sum \limits_{p n j j^\prime m m^\prime} F_{p j m; n j^\prime m^\prime} c^\dag_{p j m} c_{n j^\prime m^\prime} \\
&= \sum \limits_{p n j j^\prime m m^\prime} \langle p j m | [\boldsymbol{\sigma}_S Y_L \tau_-]_{J M} | n j^\prime m^\prime \rangle c^\dag_{p j m} c_{n j^\prime m^\prime} ,
\end{split}
\end{align}
where $\tau_-$ is the isospin lowering operator defined as $\tau_- | n \rangle = | p \rangle$ and $c^\dag_{p j m}, c_{n j^\prime m^\prime}$ are single-particle creation and annihilation operators, with angular momenta $j,j^\prime$ and its projections $m,m^\prime$.
%From above definition it follows that $\hat{F}_{np} = 0$, which corresponds to second term in Eq. (\ref{eq:W_matrix}). Similar considerations apply for the $\beta^+$ direction, except that $\hat{F}_{pn} = 0$.

\section{Numerical implementation and tests}\label{sec:numerical_implementations_and_tests}

\subsection{Numerical implementation}

Our linear response FT-PNQRPA model is developed on top of the FT-RHB model. The FT-RHB equation is solved by expanding the wave functions on the harmonic oscillator basis~\cite{Gambhir1990_APNY198-132}, and if not stated otherwise, 20 oscillator shells are used, i.e., $N_{osc}=20$.
The coordinate-space Bethe-Salpeter equation (\ref{eq:bethe_salpeter_coord_space}) for the reduced response is solved on the Gauss-Hermite radial mesh with $N_{gh}=24$ meshpoints within a box of $R = 20$ fm.
%something about pairing interaction
We note that the residual pairing interaction can be either isovector ($T=1$) or isoscalar ($T=0$) depending on the natural parity of considered excitation~\cite{BAI-CL2014_PRC90-054335,Sagawa2016_PS91-083011}. Although the form of the separable interaction remains the same (as shown in Appendix \ref{sec:app_a}), its strength changes. For natural parity transitions ($0^+,1^-$), the same interaction strength $G$ is used as in the ground-state case, while for unnatural parity transitions ($0^-,1^+,2^-$), pairing interaction strength $G$ is multiplied by the isoscalar pairing strength $V^{is}$, which is a free parameter in the model. For the $ph$ interaction we use both DD-PCX and DD-PC1 parameter sets with residual interaction channels defined in appendix \ref{sec:app_a}.
 %cut offs
  In conventional matrix QRPA calculations based on Hartree + BCS, the total number of quasiparticle pairs is restricted by the energy cut-off $E_{cut}$, otherwise the dimension of the eigenvalue problem can become too large~\cite{Paar2003_PRC67-034312,Paar2004_PRC69-054303}. Also, to remove the q.p. pairs that almost do not contribute to the strength function, a threshold on the product of the BCS occupation amplitudes $u$ and $v$ of a particular pair is set. We emphasize that in our implementation of the linear response FT-PNQRPA no such restrictions are used. We neglect the antiparticle-hole transitions~\cite{Ring2001_NPA694-249}, which is a good approximation for the charge-exchange channel~\cite{Paar2004_PRC69-054303}.
 % energy mesh
 The Bethe-Salpeter equation (\ref{eq:bethe_salpeter_coord_space}) is solved for each energy mesh point with a $\Delta E = 0.1$ MeV interval. The matrix element in Eq. (\ref{eq:contour_strength}) is calculated using a circular loop with a radius of 0.1 MeV that encloses the $i-$th pole in the response function. The contour integral is solved using the Simpson's integration rule. Similarly, the transition strength matrix elements contributed by  a particular 2 q.p. excitation (see Eq. \ref{eq:ftqrpa_excitation_me}) are calculated with the same contour around the $i-$th pole by obtaining the discrete matrix FT-QRPA eigenvectors from Eqs. (\ref{eq:eigen1}-\ref{eq:eigen4}) and calculating the matrix element of the external field operator as in matrix FT-QRPA~\cite{Yuksel2017_PRC96-024303}.  The method presented in this work is significantly faster compared to conventional matrix PNQRPA both at zero~\cite{Paar2004_PRC69-054303} and finite temperature~\cite{Yuksel2020_PRC101-044305}, making it suitable for large-scale calculations of excitation strength functions, and weak interaction processes of astrophysical relevance.

\subsection{Numerical tests}

For the numerical tests of the linear response FT-PNQRPA, we performed a comparison with the following codes based on the matrix implementation of the QRPA
\begin{itemize}
\item RHB + matrix PNQRPA  code at zero temperature based on the DD-PC1 interaction (denoted as RHB+mQRPA in the following)  \cite{deni_paper}
\item  FT-RMF (relativistic mean-field) + matrix FT-PNRPA code at finite-temperature based on the DD-PC1 interaction (denoted as FT-RMF+mFT-RPA)~\cite{NIU-YF2011_PRC83-045807}
\item FT-HBCS + matrix FT-PNQRPA code at finite-temperature based on the DD-PC1 interaction (denoted as FT-BCS+mFT-QRPA in the following)~\cite{Ravlic2020_PRC102-065804,Yuksel2020_PRC101-044305}
\end{itemize}

\begin{figure}
\centering
\includegraphics[width=\linewidth]{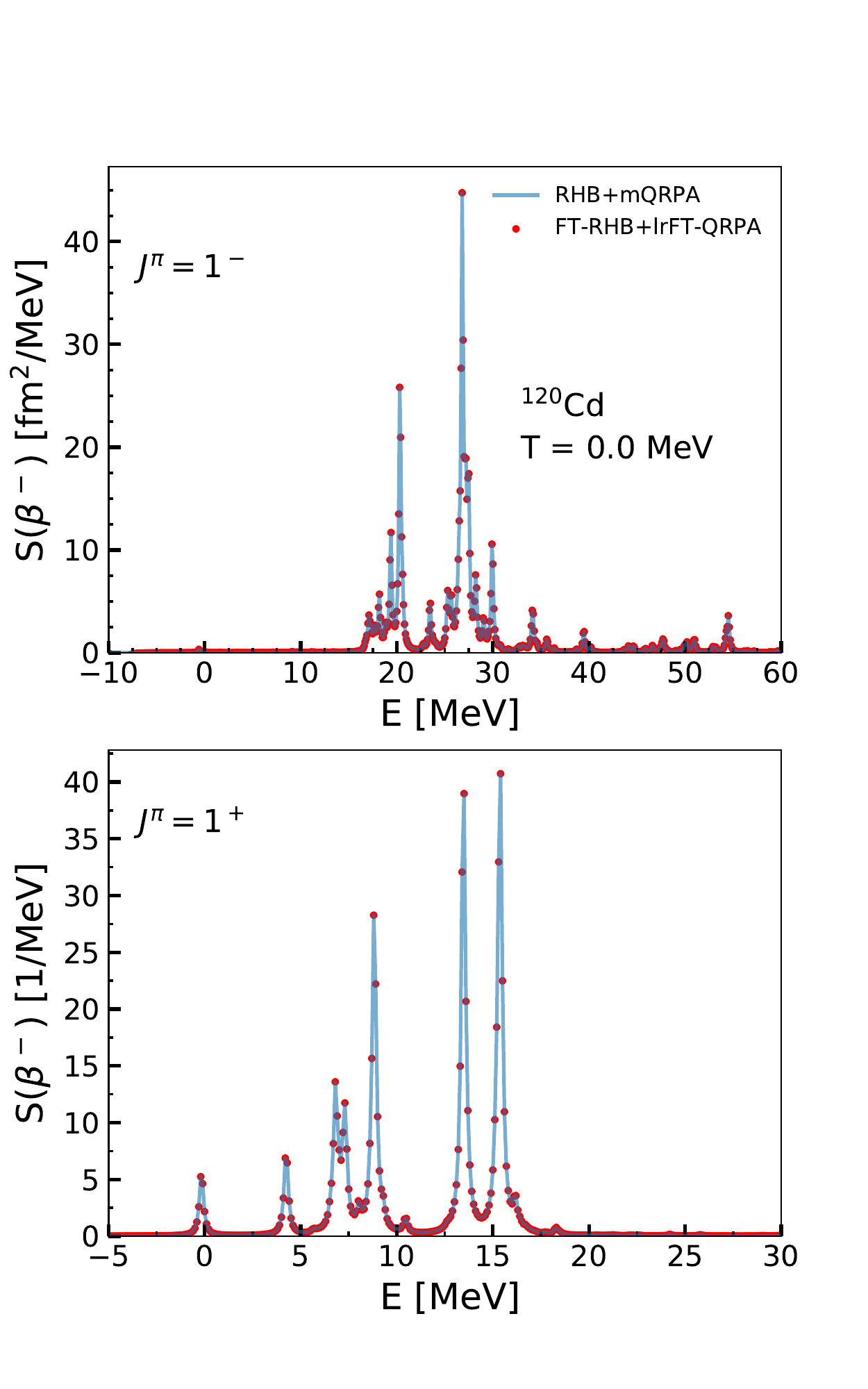}
\caption{Strength functions of $J^\pi = 1^-$ (upper panel) and $1^+$ (lower panel) excitations in $\beta^-$ direction in ${}^{120}$Cd calculated by linear response FT-PNQRPA calculation based on FT-RHB at zero-temperature (red dots), in comparison with those calculated by the matrix PNQRPA based on RHB  (blue line). In order to limit the total number of 2 q.p. pairs, $N_{osc}=12$ is used in the RHB and FT-RHB calculations.}\label{fig:comparison_rhb}
\end{figure}

In the present analysis, we set the parameter $\eta$ in Eq. (\ref{eq:reduced_unp_resp}) to 0.25 MeV in order to better visualize distinct peaks. It can be inferred that $\eta$ corresponds to smearing width parameter in the matrix QRPA defined in Refs.~\cite{Paar2004_PRC69-054303,Paar2003_PRC67-034312}. As a first check, we will compare the results of the linear response FT-PNQRPA based on the FT-RHB (denoted as FT-RHB+lrFT-QRPA) with those of RHB+mQRPA at zero temperature, in order to check the correctness of the zero-temperature limit of the linear response FT-PNQRPA code. A comparison is shown in Fig. \ref{fig:comparison_rhb} for ${}^{120}$Cd of $J^\pi = 1^-$ (upper panel) and $1^+$ (lower panel) excitations in the $\beta^-$ direction. In the RHB (or FT-RHB) code, a total of $N_{osc} = 12$ harmonic oscillator shells are used to limit the number of two-quasiparticle (2 q.p.) pairs. For the isoscalar pairing strength in the $1^+$ excitation, we use $V^{is} = 2.0$, for demonstration purposes. Our implementation of linear response FT-PNQRPA reproduces the zero-temperature limit when compared to the corresponding matrix code.

\begin{figure}
\centering
\includegraphics[width=\linewidth]{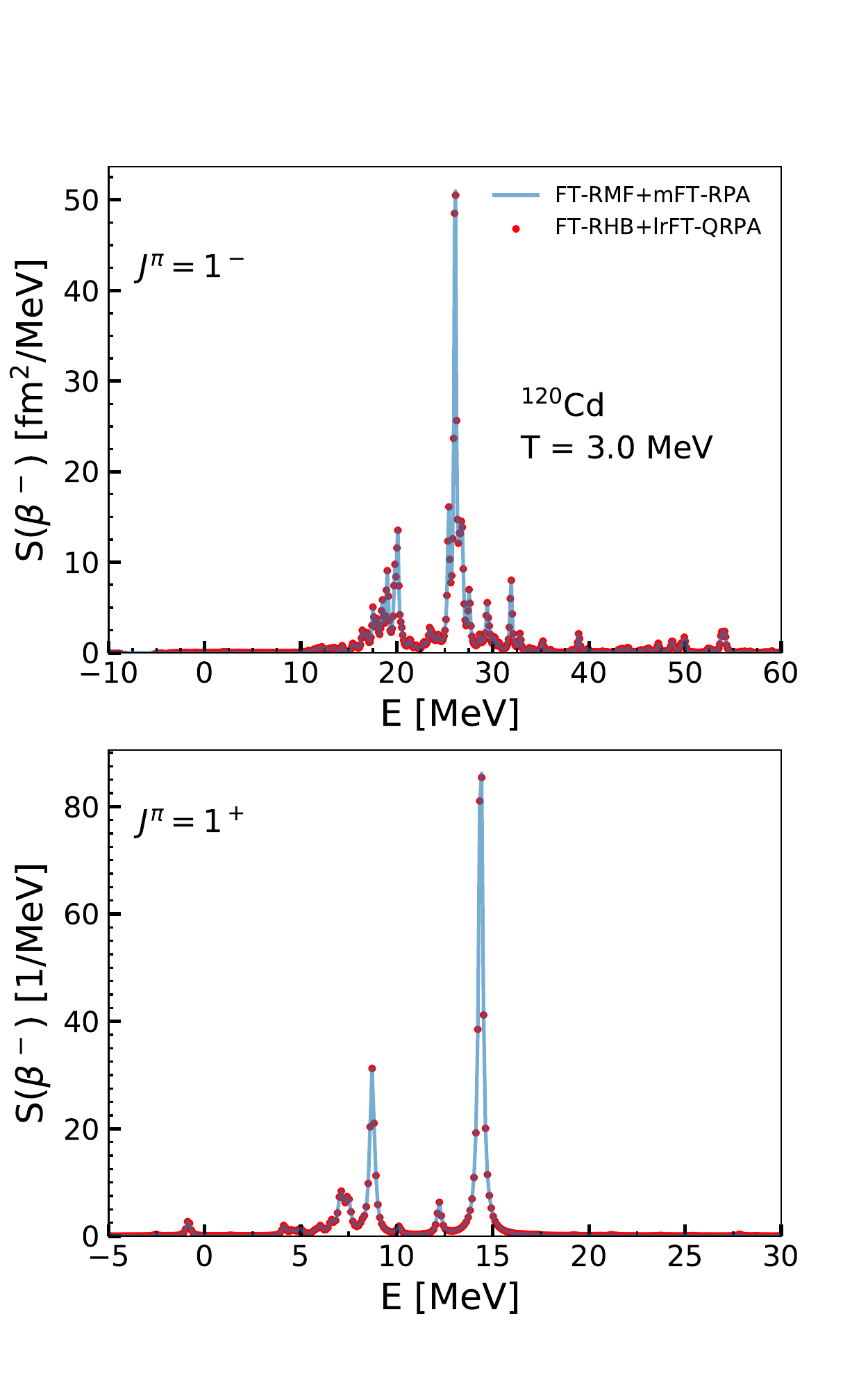}
\caption{Same as in Fig. \ref{fig:comparison_rhb} except the comparison is between the linear response FT-PNQRPA based on the FT-RHB (red dots) with the matrix FT-RPA based on the FT-RMF (blue line) at $T = 3.0$ MeV.}\label{fig:comparison_ftrpa}
\end{figure}

Next we check the implementation of temperature effects in the linear response FT-PNQRPA code, so in Fig. \ref{fig:comparison_ftrpa}, results for a relatively high temperature $T = 3$ MeV are compared between linear response FT-PNQRPA based on the FT-RHB (FT-RHB+lrFT-QRPA) and FT-PNRPA based on FT-RMF which does not include pairing correlations, again keeping $N_{osc} = 12$. Strength functions corresponding to these two calculations agree well, and this shows the correct linear response implementation in the high temperature limit where pairing correlations vanish.

\begin{figure}
\centering
\includegraphics[width=\linewidth]{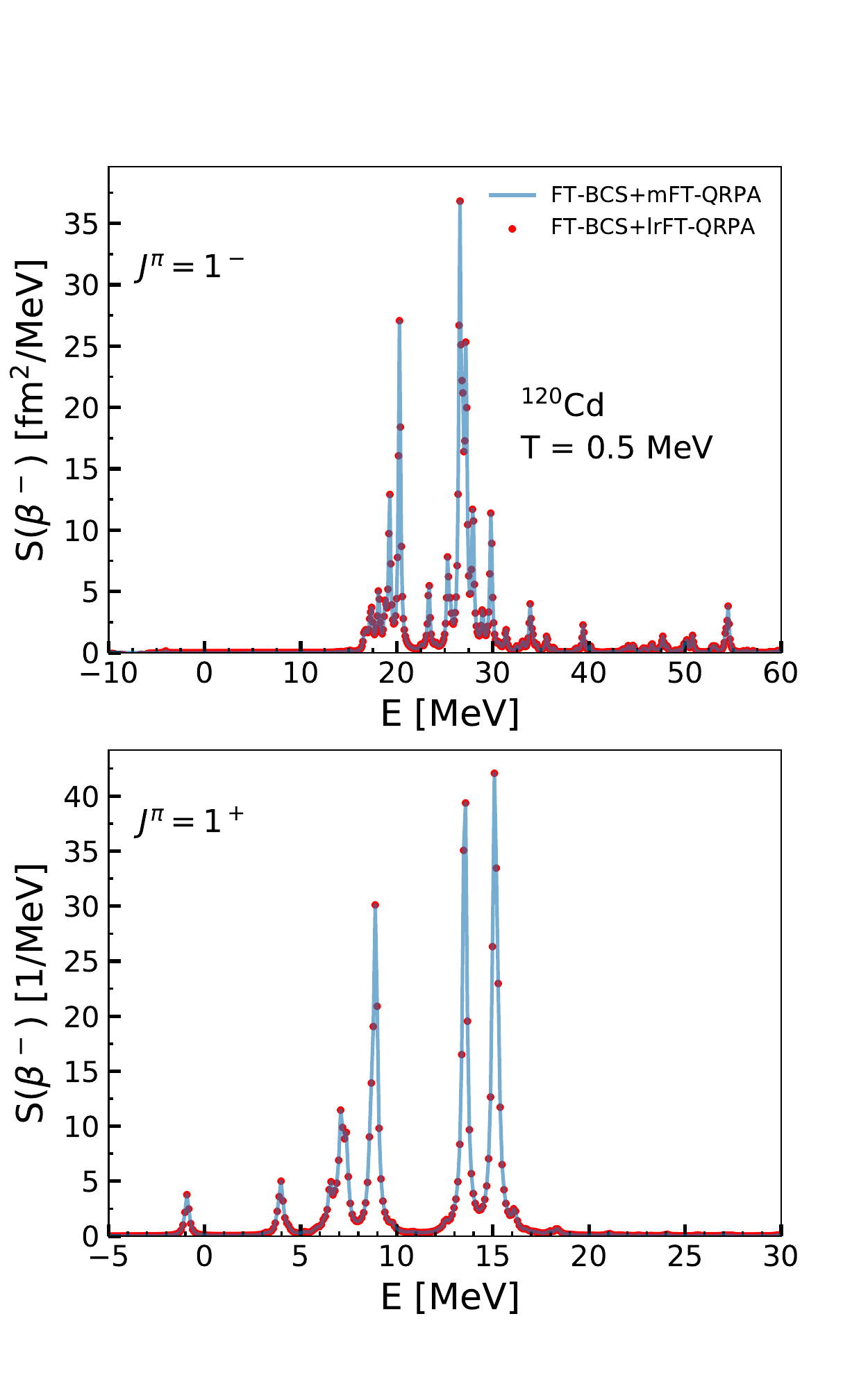}
\caption{Same as in Fig. \ref{fig:comparison_rhb} except the comparison is between the linear response FT-PNQRPA based on the FT-HBCS (red dots) with the matrix FT-PNQRPA based on the FT-HBCS (blue line) at $T = 0.5$ MeV.}\label{fig:comparison_bcs}
\end{figure}

\begin{table*}
\centering
\caption{Comparison of GT${}^-$ strength $B$(GT${}^-$) of the $1^+$ state at $E = 13.54$ MeV in ${}^{120}$Cd at $T = 0.5$ MeV, as shown in Fig. \ref{fig:comparison_bcs}, and the corresponding transition matrix elements contributed by the particular two-quasiparticle excitations $\langle i | \hat{F} | 0 \rangle_{\pi \nu} $ between lrFT-QRPA and mFT-QRPA calculations. In order to show the convergence of the method results for $N_S = 4$, $N_S = 6$ and $N_S = 8$ Simpson's integration meshpoints are shown. }\label{tab:tab_comp}
\begin{tabular}{cc|ccc}
\hline
\hline
 & mFT-QRPA &   & lrFT-QRPA & \\
 \cline{3-5}
 &           & $N_S=4$ & $N_S=6$ & $N_S = 8$ \\
 \hline
B(GT${}^-$) & 16.828564095 & \textbf{16.8}32567631 & \textbf{16.828}851131 & \textbf{16.8285}72685\\
\hline
Transition $i$ & $\langle i | \hat{F} | 0 \rangle_{\pi \nu} $ & $\langle i | \hat{F} | 0 \rangle_{\pi \nu}$ & $\langle i | \hat{F} | 0 \rangle_{\pi \nu}$ & $\langle i | \hat{F} | 0 \rangle_{\pi \nu}$ \\
\hline
$(\nu 3s_{1/2}, \pi 3s_{1/2})$ & 0.111059165 & \textbf{0.1110}86451 & \textbf{0.111059}201 & \textbf{0.111059}374\\
$(\nu 2d_{5/2}, \pi 2d_{3/2})$ & 0.345657488 & \textbf{0.345}729591 & \textbf{0.34565}8125 & \textbf{0.345657}672\\
$(\nu 2d_{3/2}, \pi 2d_{5/2})$ & 0.163268438 & \textbf{0.1632}99678 & \textbf{0.163268}732 & \textbf{0.163268}551\\
$(\nu 2d_{5/2}, \pi 2d_{5/2})$ & 0.193705886 & \textbf{0.1937}49887 & \textbf{0.19370}6236 & \textbf{0.19370}6013\\
$(\nu 1g_{7/2}, \pi 1g_{7/2})$ & 0.156681034 & \textbf{0.156}709583 & \textbf{0.156681}123 & \textbf{0.1566810}87\\
$(\nu 1g_{9/2}, \pi 1g_{7/2})$ & 3.170115103 & \textbf{3.1}69786542 & \textbf{3.17011}4949 & \textbf{3.170115}424\\
$(\nu 1h_{11/2}, \pi 1h_{9/2})$ & -0.373797786 & \textbf{-0.373}204264 & \textbf{-0.3737}62528 & \textbf{-0.37379}8104\\
\hline
\end{tabular}
\end{table*}

In order to test the more general case with both the pairing correlations and temperature effects present, we have constructed the linear response FT-PNQRPA on top of the FT-HBCS ground-state (denoted as FT-BCS+lrFT-QRPA), and compared it with the FT-HBCS + matrix FT-PNQRPA code. The FT-HBCS code employs the delta-pairing force as defined in Eq. (6) of Ref.~\cite{Bender2000_EPJA8-59} with strength $V_{0,p} = V_{0,n} = -300$ MeV fm${}^3$ for both protons and neutrons. We note that for this test both linear response FT-PNQRPA and matrix FT-PNQRPA use the same form of delta-pairing in the FT-HBCS ground-state, while at the level of the residual interaction separable pairing is used. Results for the ${}^{120}$Cd at $T = 0.5$ MeV are shown in Fig. \ref{fig:comparison_bcs}, where we take $N_{osc} = 12$ harmonic oscillator shells, to limit the size of FT-PNQRPA matrix for comparison and isoscalar pairing strength in $1^+$ excitation is set to $V^{is} = 2.0$, again for demonstration. We note that in this case both temperature and pairing effects are present, with pairing also included in the residual interaction, which displays the most general case studied within this work. Agreement between the two codes is excellent.

An additional test can be made by explicitly calculating the matrix FT-PNQRPA eigenmodes and corresponding transition matrix elements using the linear response formalism detailed in Appendix \ref{sec:appc}. For this test, we use the previous calculation for the $J^\pi = 1^+$ excitation at $T = 0.5$ MeV in ${}^{120}$Cd. We select the peak at $E = 13.54$ MeV and enclose it with a circular contour of 0.05 MeV radius in the complex energy plane. The matrix FT-PNQRPA eigenvectors are then calculated by solving the contour integrals in Eqs. (\ref{eq:eigen1}-\ref{eq:eigen4}) using Simpson's rule with $N_S$ integration mesh points, while the corresponding transition strength is calculated with Eq. (\ref{eq:contour_strength}) using the same integration meshes. Results for the transition strength of the $J^\pi = 1^+$ state at $E=13.54$ MeV and the transition matrix elements from the selected 2 q.p. pairs with largest contribution are shown in Tab. \ref{tab:tab_comp} for $N_S = 4$, $N_S = 6$ and $N_S = 8$.  Results for $N_S = 6$ already show a good convergence, having agreement up to 5 or more significant digits. Improvement when going from $N_S = 6$ to $N_S = 8$ is only up to one significant digit, leading to the conclusion that the optimal number of Simpson's meshpoints for the contour integration is $N_S = 6$ due to faster execution time.

%The above tests ensures the correct implementation of linear response FT-PNQRPA code based on FT-RHB.

\section{Illustrative calculations in tin isotopic chain}\label{sec:illustrative_calculations}

\begin{figure}
\centering
\includegraphics[width=\linewidth]{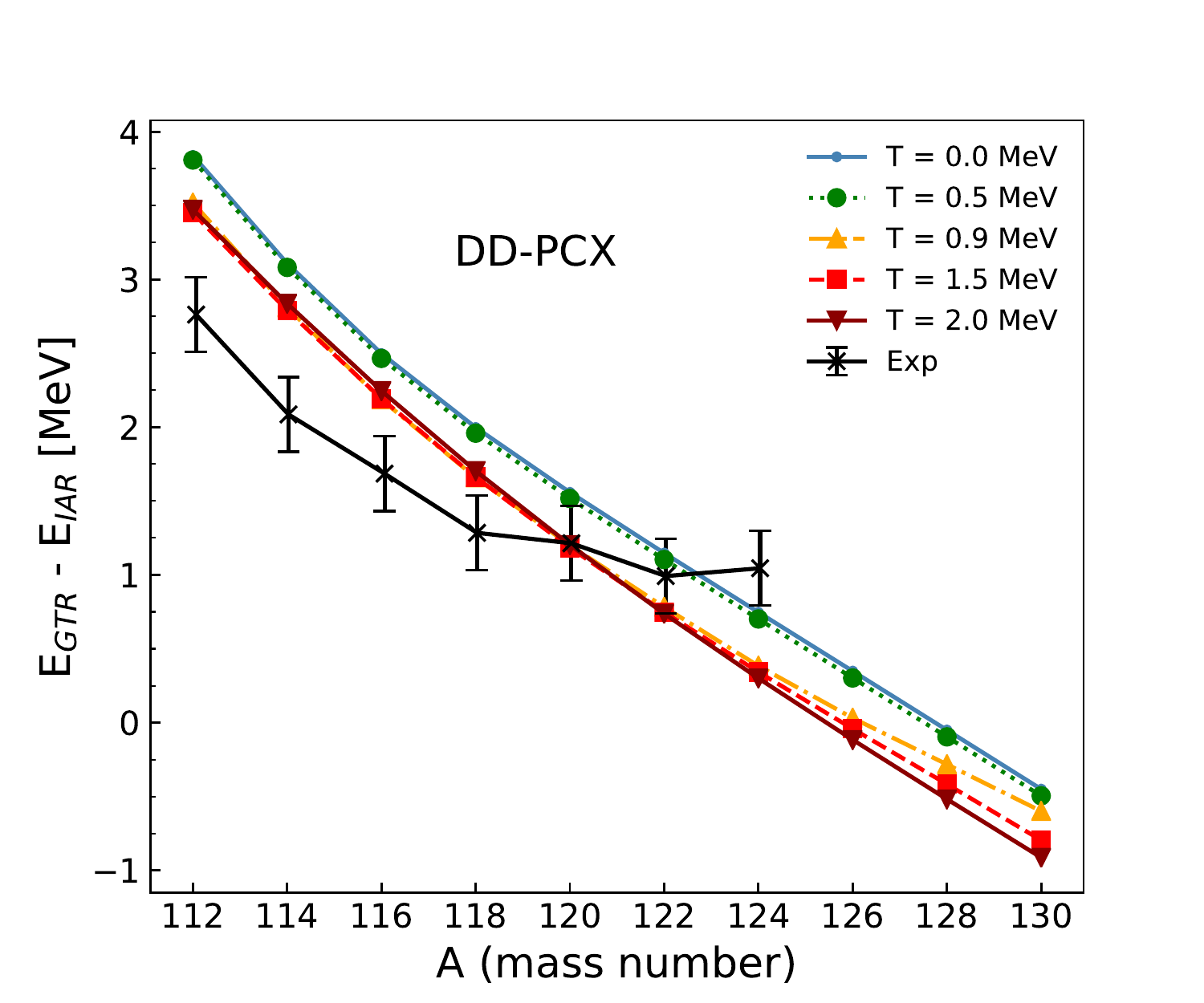}
\caption{Difference between centroid energies of Gamow-Teller and Isobaric Analog Resonances $E_{\rm GTR}-E_{\rm IAR}$ for even-even tin isotopes as a function of  mass number $A=112-130$ at temperatures T = $0,0.5,0.9,1.5$ and 2.0 MeV. Experimental data are taken from Ref.~\cite{Pham1995_PRC51-526}.  }\label{fig:skin}
\end{figure}

In this section, we present the calculations of various spin-isospin excitations at different temperatures by linear response FT-PNQRPA based on the FT-RHB model. We choose even-even tin isotopes in the range $A=112-134$ representing open-shell nuclei where the pairing interaction is present only for neutrons in the ground state due to the $Z = 50$ shell closure. We will fix the isoscalar pairing strength in unnatural-parity transitions to $V^{is}=1.5$, guided by the study of the difference between GTR and IAR centroid energies for relativistic point-coupling functionals within this work (see Fig. \ref{fig:skin}) and in Ref.~\cite{Vale2021_PRC103-064307}. We note that for natural-parity transitions only isovector pairing $(T=1)$ is present and determined in the ground-state FT-RHB calculation. To explore the temperature effects on charge-exchange excitations, we study the centroid energy evolution, defined as
\begin{equation}\label{eq:centroid}
E_{cent.} = \frac{m_1}{m_0},
\end{equation}
where the $k-$th moment is defined as $m_k = \int d\omega \omega^k S_F(\omega)$. %\textcolor{red}{The summed strength up to excitation energy $E$ can be calculated from the $m_0$ as \cite{ring2004nuclear}
%\begin{equation}\label{eq:m_0_str}
%\sum \limits_{E_i < E} |\langle i | \hat{F} | 0 \rangle|^2  = -\frac{1}{\pi} %\int \limits^E d \omega S_F (\omega),
%\end{equation} where $|\langle i | \hat{F} | 0 \rangle|^2$ is the matrix FT-QRPA strength and $E_i$ is the corresponding eigenvalue. In the following analysis, we will abbreviate $|\langle i | \hat{F} | 0 \rangle|^2$ as $B_i(J^\pi)$.}
Using this approach, we investigate the most general case where both pairing and temperature effects are present. For all calculations, the number of oscillator shells in the ground-state calculation is $N_{osc}=20$, and no additional constraint on 2 q.p. pairs is set. This demonstrates the computational efficiency of the linear response QRPA calculation compared to the conventional matrix QRPA. For example, for $J^\pi = 2^-$ excitations with $N_{osc}=20$ and no additional cut-off on 2 q.p. pairs, total number of pairs is close to 7500 which results in a dimension of $30000 \times 30000$ for the QRPA matrix at finite-temperature using the conventional matrix approach as in Refs.~\cite{Yuksel2020_PRC101-044305,Yuksel2017_PRC96-024303,Ravlic2020_PRC102-065804}, and the diagonalization of such a big matrix in the matrix QRPA is very time-consuming, while the present linear response QRPA approach avoids such a diagonalization problem. The smearing width $\eta$ is set to 1 MeV in accordance with QRPA calculations in Ref.~\cite{Paar2004_PRC69-054303}.

We limit our study to the Fermi ($J^\pi = 0^+$), Gamow-Teller ($J^\pi = 1^+$) and spin-dipole ($J^\pi = 0^-, 1^-, 2^-$) excitations. The excitation strength is studied within the temperature interval of $T = 0$ to $T = 1.5$ MeV. Due to the grand-canonical treatment of the nuclear ground state at finite temperature, a sharp phase transition is obtained at the critical temperature $T_c$ where pairing correlations vanish~\cite{Goodman1981_NPA352-30}. In Tab. \ref{tab:critical_temp} the neutron critical temperatures $T_c^n$ together with mean pairing gaps at zero temperature $\Delta^0_n$ are shown for selected tin isotopes calculated both with DD-PC1 and DD-PCX interactions. The separable pairing interaction with the parameterization described in Sec. \ref{sec:theoretical_formalism} is used. It is observed that $T_c^n$ is higher for the DD-PCX interaction because of the larger pairing strength parameters (cf. Sec. \ref{sec:theoretical_formalism} and Ref.~\cite{Yuksel2019_PRC99-034318}). From table \ref{tab:critical_temp} it follows that we can neglect the pairing correlations for the considered tin nuclei above $T \approx 0.8(1.0)$ MeV for DD-PC1(DD-PCX).

\begin{table}[htb]
\centering
\caption{The neutron critical temperature $T_c^n$ and mean pairing gap $\Delta_n^0$ at zero temperature for particular even-even tin isotopes considered within this work. Results are calculated with the DD-PC1 and DD-PCX interactions.}\label{tab:critical_temp}
\begin{tabular}{ccc|cc}
\hline
 & \multicolumn{2}{c|}{DD-PC1} & \multicolumn{2}{c}{DD-PCX}  \\
 \hline
nucleus & $T_c^n$ [MeV] & $\Delta_n^0$ [MeV] & $T_c^n$ [MeV] & $\Delta_n^0$ [MeV]   \\
\hline
\hline
${}^{112}$Sn & 0.81 & 1.31 & 1.06 & 1.73 \\
${}^{116}$Sn & 0.79 & 1.25 & 1.04 & 1.65 \\
${}^{120}$Sn & 0.80 & 1.34 & 1.00 & 1.64\\
${}^{124}$Sn & 0.76 & 1.31 & 0.93 & 1.56\\
${}^{128}$Sn & 0.66 & 1.08 & 0.80 & 1.28 \\
\hline
\end{tabular}
\end{table}

\begin{figure*}
\centering
\includegraphics[width=\linewidth]{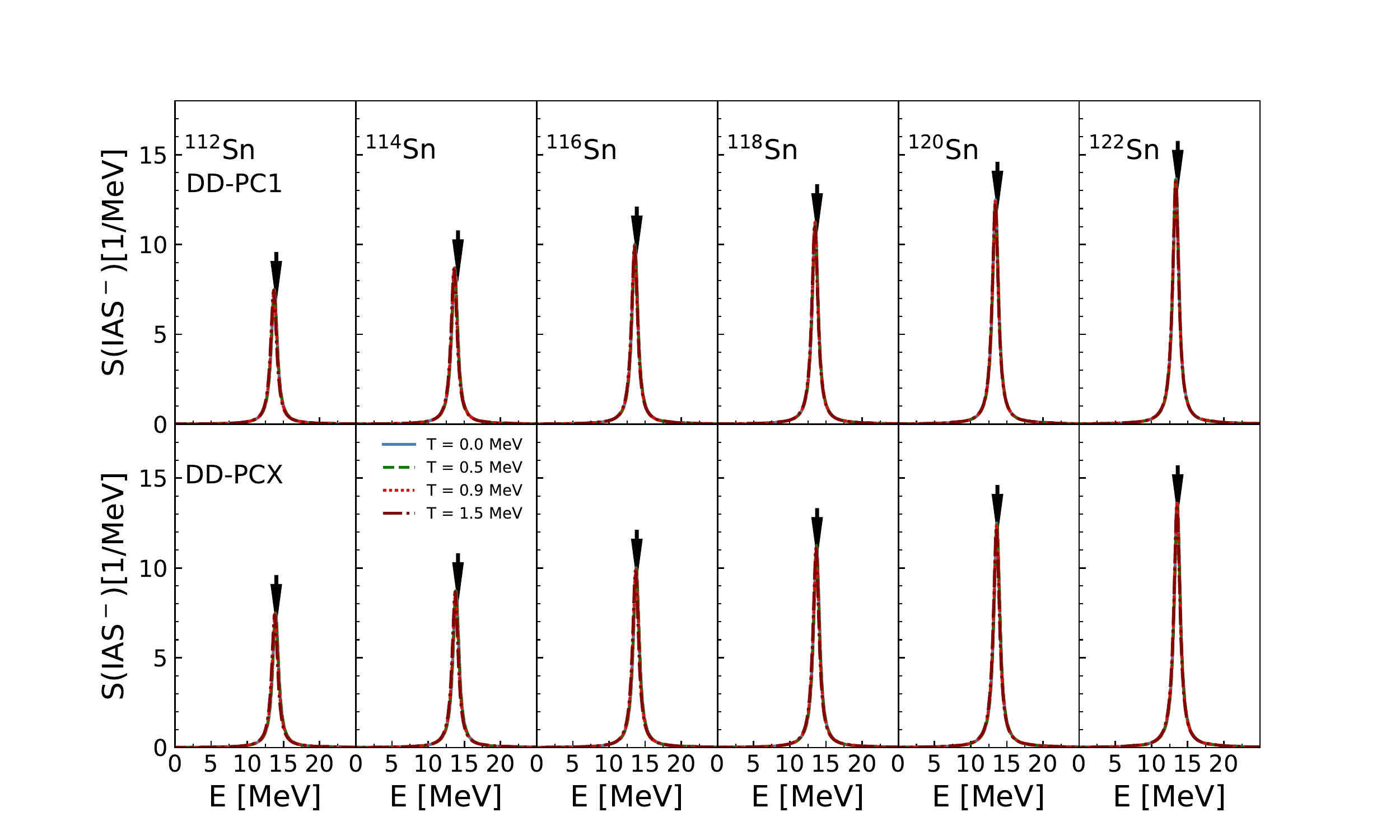}
\caption{The $J^\pi = 0^+$ strength functions in $A=112-122$ even-even tin isotopes with respect to the excitation energy of the parent nucleus for temperatures $T = 0, 0.5, 0.9$ and 1.5 MeV calculated by linear response FT-PNQRPA using the DD-PC1 (upper panel) and DD-PCX (lower panel) interaction. Black arrows denote the experimental centroid energies from Ref. \cite{Pham1995_PRC51-526}. }\label{fig:ias}
\end{figure*}

%ias
The Fermi strength distribution ($J^\pi = 0^+$) is obtained by setting $S = 0$ and $L = 0$ in the matrix element of Eq. (\ref{eq:external_matrix_element}) and solving the reduced Bethe-Salpeter equation for the linear response strength function. Since this is a natural-parity transition, the residual pairing interaction is an isovector one, and we adopt the same pairing strengths for protons and neutrons as for the ground-state calculation with DD-PC1, while their average is taken for DD-PCX (see appendix \ref{sec:app_a} for details). The temperature evolution of the Fermi excitation strengths calculated with DD-PC1 (upper panel) and DD-PCX (lower panel) interactions for even-even $A=112-122$ tin isotopes are shown in Fig. \ref{fig:ias}. Experimental centroid energies from Ref.~\cite{Pham1995_PRC51-526} are denoted with black arrows. Our results agree with experimental data within the interval of 1 MeV for both considered functionals, with better agreement for DD-PCX. The importance of self-consistent calculations is especially exemplified for the Fermi strength function, as noted in Ref.~\cite{Paar2004_PRC69-054303}. A common test of self-consistency is to neglect the Coulomb interaction in the FT-RHB calculation at zero-temperature. In that case the nuclear Hamiltonian commutes with the isospin operator, representing a good isospin symmetry. Therefore, the IAR strength should be located at zero energy with respect to the parent nucleus and have a strength corresponding to $N-Z$~\cite{Ring1980}. We have verified that our implementation satisfies this test, and thus shows the self-consistency of our model. If the interaction is not implemented self-consistently, the strength function would be fragmented as described in Ref.~\cite{Paar2004_PRC69-054303}. As the number of neutrons increases, the IAR strength also increases, while the IAR centroid energy shifts from 13.60(13.78) MeV to 13.27(13.45) MeV when going from ${}^{112}$Sn to ${}^{122}$Sn at zero temperature for the DD-PC1(DD-PCX) interaction. From Fig. \ref{fig:ias}, it can be observed that the temperature almost does not have an effect on the IAR strength and excitation energy. This is because the Coulomb energy difference between parent and daughter nuclei remains stable within the considered temperature interval. Therefore, since the Coulomb energy difference corresponds to IAR excitation energy, it also displays only minor changes at finite-temperature.

%From Fig. \ref{fig:ias} it can be observed that temperature does not have a significant effect on the IAR strength. With the DD-PC1 interaction at zero-temperature IAR peak is located at E = 13.3 MeV while at T = 1.5 MeV it is located at E = 12.8 MeV. This slight shift can be also observed in the insets in Fig. \ref{fig:ias} wich show temperature evolution of the centroid energy as defined in Eq. (\ref{eq:centroid}). Similar conclusions follow for DD-PCX interaction as well.

\begin{figure}[htb]
\centering
\includegraphics[width=\linewidth]{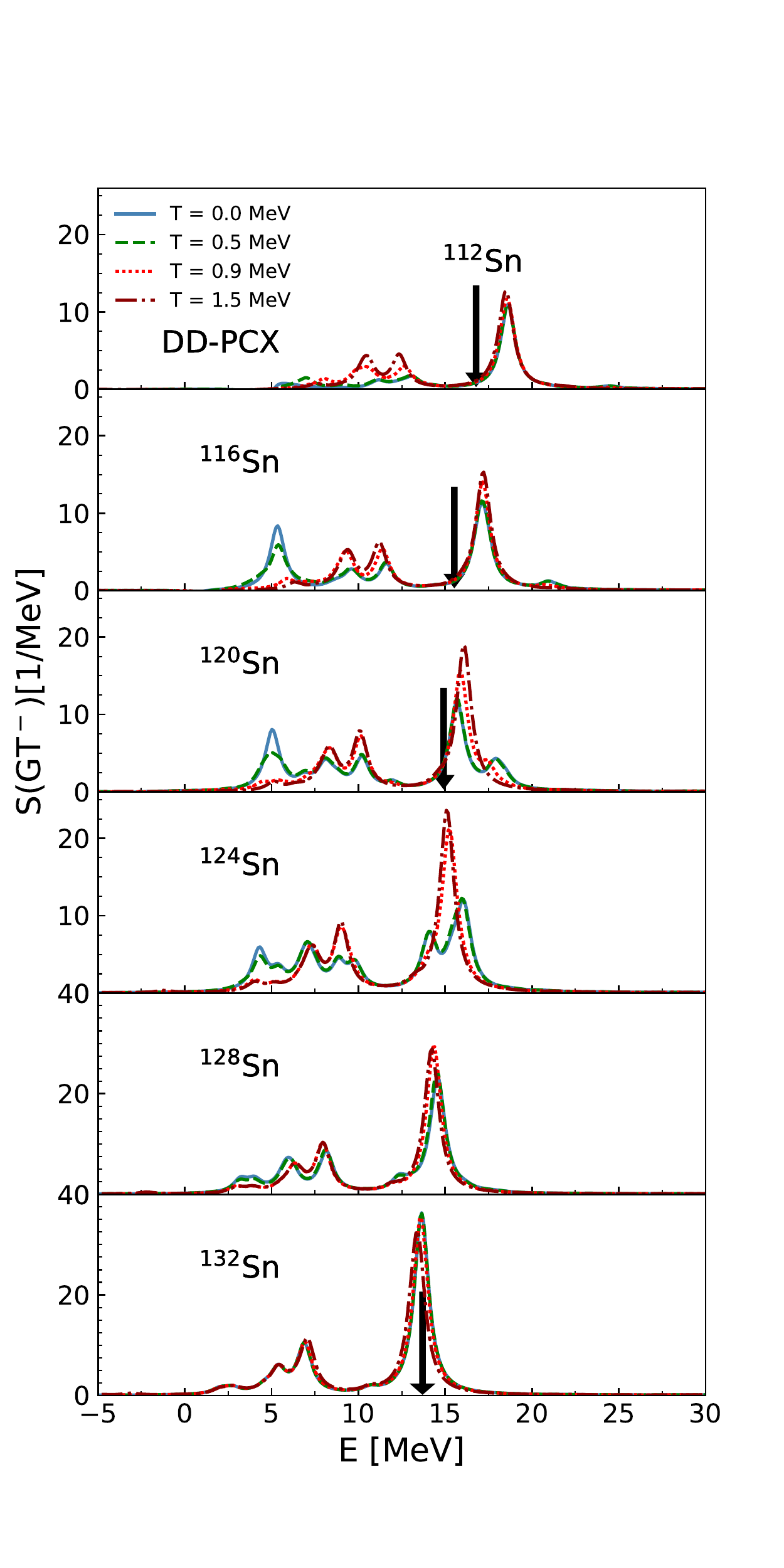}
\caption{The evolution of the Gamow-Teller ($J^\pi = 1^+$) strength function with temperature for selected even-even tin nuclei with respect to the excitation energy of the parent nucleus for temperatures $T = 0, 0.5, 0.9$ and 1.5 MeV calculated by linear response FT-PNQRPA model with DD-PCX interaction. Black arrows denote the experimental centroid energies from Refs.~\cite{Pham1995_PRC51-526,Yasuda2018_PRL121-132501}.  }\label{fig:gt}
\end{figure}

\begin{figure}[htb]
\centering
\includegraphics[width=\linewidth]{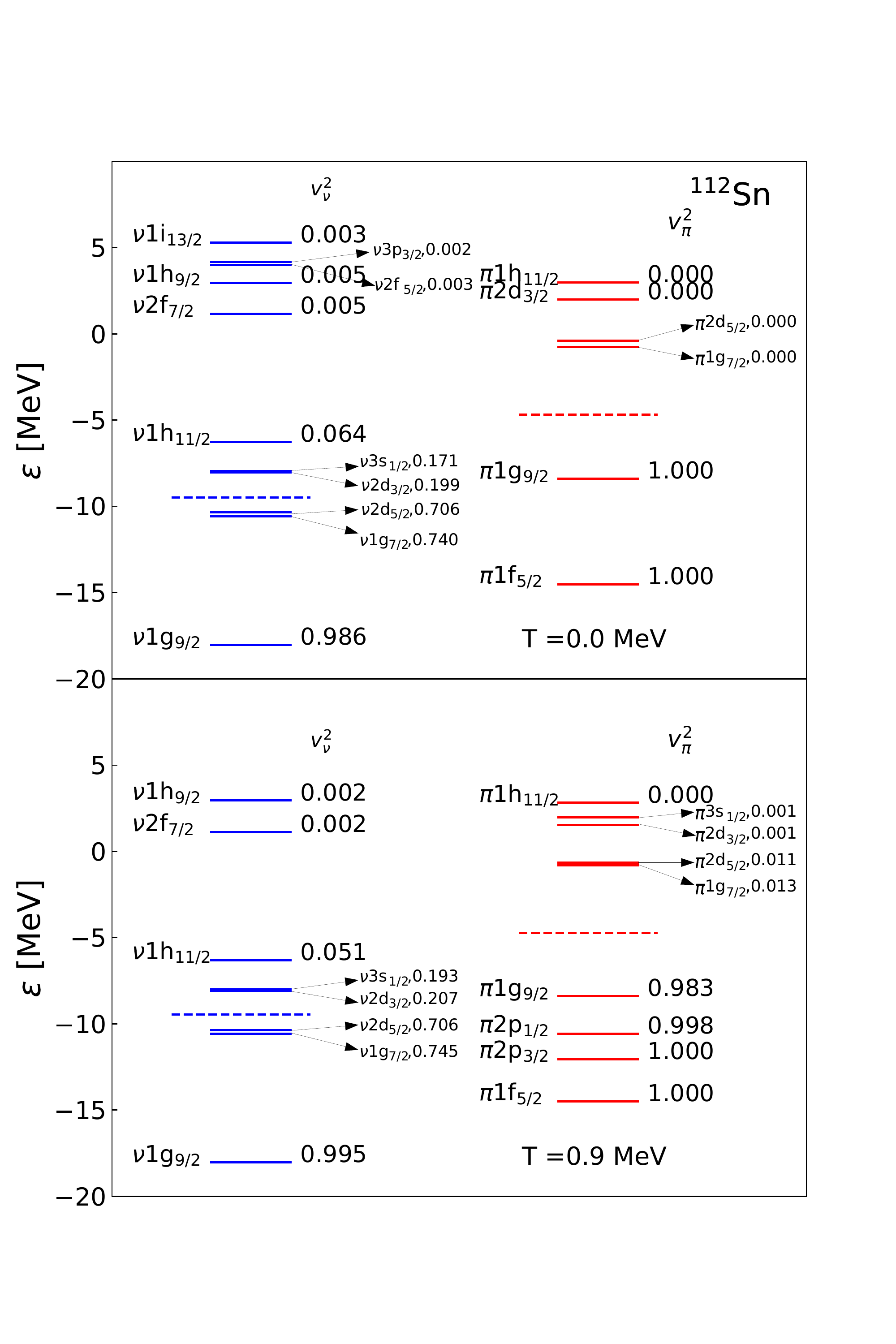}
\caption{The single-particle energy levels in canonical basis for neutrons (blue) and protons (red) calculated by FT-RHB at $T = 0$ (upper panel) and $T = 0.9$ MeV (lower panel) using the canonical transformation to obtain single-particle energies $\varepsilon$ and occupation probabilities $v_{\nu(\pi)}^2$ for neutrons(protons). Dashed lines denote the Fermi levels. }\label{fig:qplevels}
\end{figure}

%gamow-teller
In the following we limit our discussion to the DD-PCX interaction. Similar conclusions also follow for DD-PC1. The temperature evolution of Gamow-Teller ($J^\pi = 1^+$) strength function is shown for ${}^{112-132}$Sn in Fig. \ref{fig:gt}. It can be observed that the Gamow-Teller strength is split into two main peaks: (i) low-lying peaks mainly composed of the core-polarization $(\nu j = l \pm 1/2, \pi j = l \pm 1/2)$ and back spin-flip $(\nu j = l -1/2, \pi j = l +1/2)$ 2 q.p. excitations and (ii) the main GTR peak at higher excitation energies composed mainly of direct spin-flip 2 q.p. excitations $(\nu j = l+1/2, \pi j = l-1/2)$. With increasing neutron number, the overall strength function shifts to lower excitation energies, while the total strength in GTR increases, a trend observed for all considered temperatures. The excitation energy of the main peak of the doubly-magic ${}^{132}$Sn is in excellent agreement with experimental centroid energy from Ref.~\cite{Yasuda2018_PRL121-132501} indicated by black arrow in Fig. \ref{fig:gt}. Although the main peaks of the ${}^{112}$Sn, ${}^{116}$Sn, and ${}^{120}$Sn display a difference in the excitation energy when compared to experimental data from Ref.~\cite{Pham1995_PRC51-526}, these differences are at most around 2 MeV. We note that the QRPA considers only 2 q.p. excitations, while the inclusion of higher-order terms via the particle-vibration coupling (PVC) should improve the agreement~\cite{NIU-YF2012_PRC85-034314,NIU-YF2018_PLB780-325}. Furthermore, adjusting the isoscalar pairing strength for individual nucleus can slightly improve the difference, however, most studies prefer global fits to a particular functional form~\cite{Marketin2016_PRC93-025805,Niu2013_PLB723-172}. It is also observed that instead of having one prominent GTR peak, the strength can be fragmented as visible for ${}^{120}$Sn and ${}^{124}$Sn, where the main GTR peak is split into two peaks. A similar splitting was described in Refs.~\cite{Paar2004_PRC69-054303,Vale2021_PRC103-064307}, where its evolution with respect to the isoscalar pairing strength at zero temperature was studied. This fragmentation disappears for high enough temperatures where pairing correlations vanish. The temperature can influence both the low-lying and GTR strength with effects being visible already at $T = 0.5$ MeV. In order to explain the temperature evolution, we need to study the particular nuclear structure properties of selected nuclei.

\begin{figure}[htb]
\centering
\includegraphics[width=\linewidth]{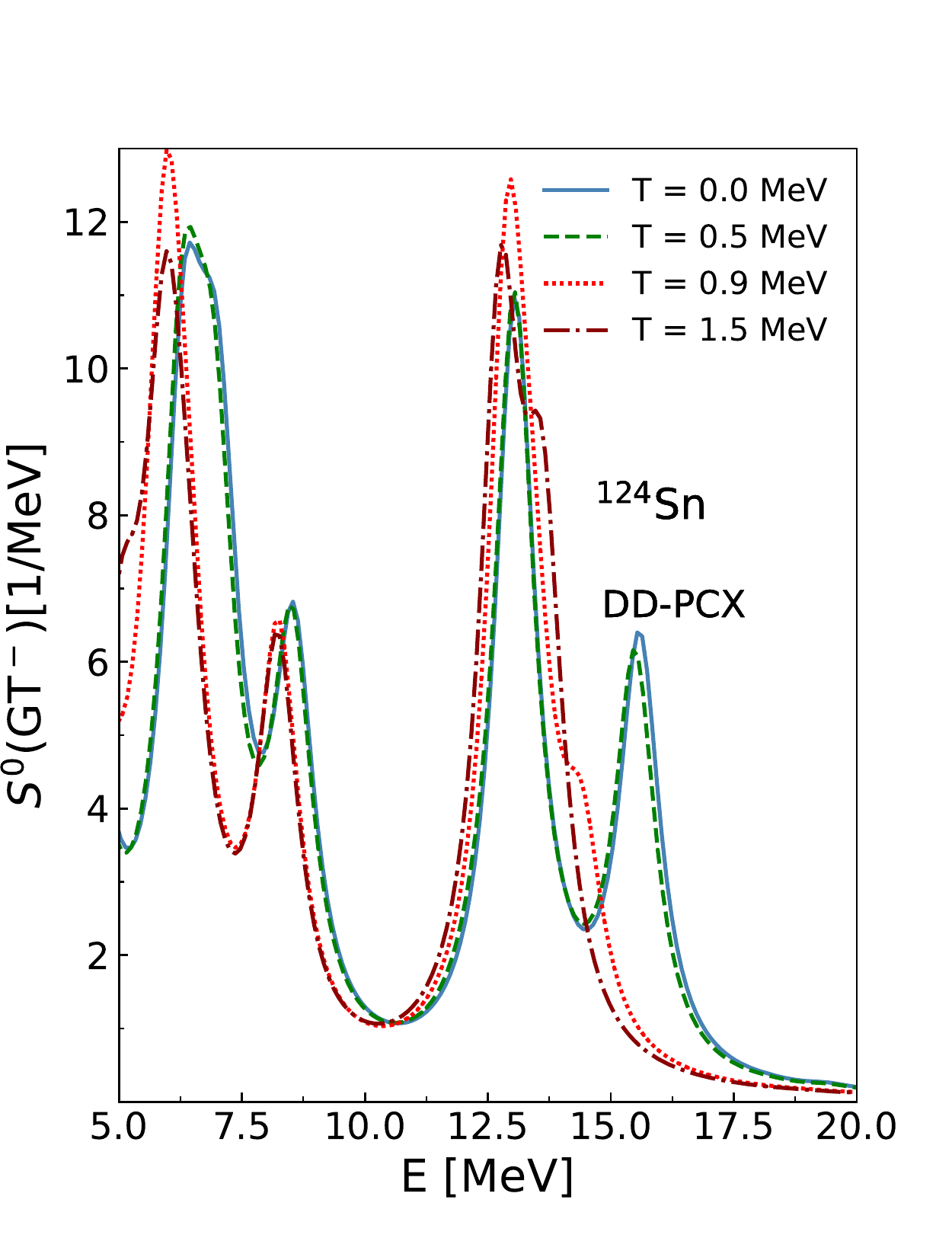}
\caption{Temperature evolution of the unperturbed $J^\pi = 1^+$ strength distributions in ${}^{124}$Sn for $T = 0, 0.5, 0.9$ and 1.5 MeV calculated with DD-PCX interaction. }\label{fig:gt_unp}
\end{figure}

%sn112

We start our discussion for ${}^{112}$Sn at zero-temperature. From Fig. \ref{fig:gt}, it is seen that almost the entire strength is concentrated in the GTR peak at $E = 18.62$ MeV. In order to better explain the structure of 2 q.p. pairs contributing to the GTR, in Fig. \ref{fig:qplevels} we show the single-particle levels calculated by FT-RHB  in the canonical basis for neutrons  and protons  at $T = 0$ and $T = 0.9$ MeV.  The single-particle energies $\varepsilon$ and occupation probabilities $v^2$ are obtained by performing the canonical transformation of Bogoliubov quasi-particle states as described in Ref.~\cite{Ring1980}. At zero temperature, most contributions to the GTR originate from the $(\nu 1g_{9/2}, \pi 1g_{7/2})$ transition, and some are from the $(\nu 2d_{5/2}, \pi 2d_{3/2})$ transition. At $T = 0.9$ MeV, the occupation probability of the $\nu 1g_{9/2}$ transition increases due to the weakening of the pairing correlations by the temperature effect, and thus the contribution of the $(\nu 1g_{9/2}, \pi 1g_{7/2})$ transition to the GTR strength is increased.  From Fig. \ref{fig:qplevels}, it is observed that a higher temperature leads to the unblocking of some proton q.p. levels, e.g., $\pi 1g_{9/2}$. These temperature effects combined with the vanishing of pairing correlations lead to an enhancement of the low-lying strength, where among many other 2 q.p. excitations, $(\nu 2d_{5/2}, \pi 2d_{3/2})$ dominates.

By examining the strength function of ${}^{116}$Sn, a significant change in the low-lying strength can be noticed. Compared to ${}^{112}$Sn, a strong peak appears at  $E = 5.34$ MeV at zero temperature. It is noticed that with increasing temperature, the strength of the peak decreases, diminishing completely for temperatures above the pairing collapse. The peak stems from 4 additional neutrons in ${}^{116}$Sn, and is dominated by the $pp$ part of the residual interaction in the $(\nu 2d_{3/2}, \pi 2d_{5/2})$ transition, with also significant contributions from core-polarization transitions. The GTR peak at zero temperature, now located at $E = 17.12$ MeV, is still dominated by the $(\nu 1g_{9/2}, \pi 1g_{7/2})$ transition, yet with increasing mixing of the $(\nu 1h_{11/2}, \pi 1h_{9/2})$ transition due to the strong $pp$ interaction.

By further increasing the neutron number, a fragmentation of GTR strength occurs in ${}^{120}$Sn and ${}^{124}$Sn at zero temperature. To describe the splitting of the GTR, we show the unperturbed strength function of ${}^{124}$Sn in Fig. \ref{fig:gt_unp}. The unperturbed GT response peaks correspond to singularities of unperturbed response function defined in Eq. (\ref{eq:reduced_unp_resp}). At zero temperature, two peaks are of interest in the unperturbed response: (i) the state at $E = 13.06$ MeV corresponding to the $(\nu 1g_{9/2}, \pi 1g_{7/2})$ 2 q.p. transition, (ii) the state at $E = 15.58$ MeV corresponding to the $(\nu 1h_{11/2}, \pi 1h_{9/2})$ transition. The total unperturbed excitation energy is just a sum of proton and neutron quasiparticle energies $E = E_\pi + E_\nu$, where quasiparticle energies are determined from the FT-RHB ground-state calculations. Once we include the residual interaction, the GTR splitting originates from these two peaks. From Fig. \ref{fig:gt_unp}, it is observed that as the temperature increases, the unperturbed energy difference between the described two transitions reduces. Above the pairing-collapse temperature, these two peaks become nearly degenerate in the unperturbed energy thus coherently contributing to the GTR, once the residual interaction is included in the calculation.
%Here we stress the importance of the self-consistent finite-temperature calculations and its influence on the GTR. For temperatures above the pairing collapse temperature evolution follows simple FT-RRPA calculations as described previously.

%\begin{figure}
%\centering
%\includegraphics[width=\linewidth]{figs/gtr_strength_ll_hl.pdf}
%\caption{\textcolor{red}{Summed strength of the Gamow-Teller ($J^\pi=1^+$) response function with respect to the neutron number $N$ of tin nuclei at temperatures $T=0,0.5,0.9$ and 1.5 MeV for (a) the Gamow-Teller resonance (GTR) region and (b) the low-lying strength. Calculations are performed with the DD-PCX interaction.}}\label{fig:summed_gtr_strength}
%\end{figure}

%now describe breaking of the degeneracy in Sn128 and finish with Sn132
For ${}^{128}$Sn, the GTR has again one prominent peak at zero temperature at $E = 14.56$ MeV. Due to increased neutron number, the strength of the $(\nu 1h_{11/2}, \pi 1h_{9/2})$ transition increases, and the difference between the unperturbed energies described previously in Fig. \ref{fig:gt_unp} decreases, thus combining the peaks into the GTR by the residual interaction. With increasing temperature, the strength of the main peak slightly increases and it shifts to lower excitation energies due to the weakening of the pairing interaction. Finally, for the doubly-magic nucleus ${}^{132}$Sn, the temperature evolution of the GT strength is solely determined by the finite-temperature effects. The main GTR peak at $T = 0 $ MeV is found at $E = 13.64$ MeV, mainly dominated by $(\nu 1h_{11/2}, \pi 1h_{9/2})$ transition. With increasing temperature, the strength of the main peak decreases due to the softening of the residual interaction.

\begin{figure*}
\centering
\includegraphics[width=\linewidth]{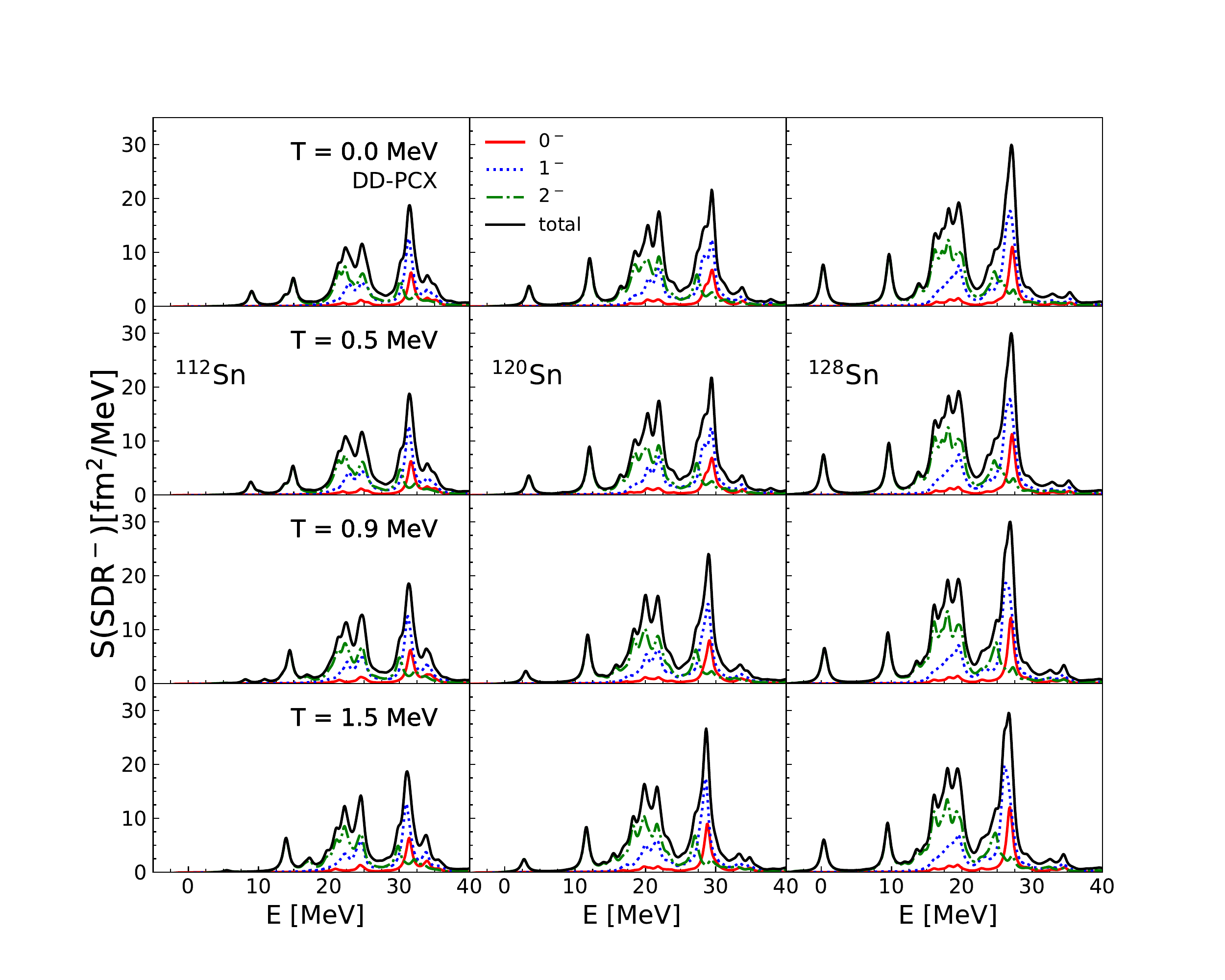}
\caption{The evolution of the spin-dipole excitation strength with temperature in ${}^{112}$Sn (left panel), ${}^{120}$Sn (center panel) and ${}^{128}$Sn (right panel) for $T = 0, 0.5, 0.9$ and 1.5 MeV calculated using the DD-PCX interaction. On the same figure are shown $J^\pi = 0^-$ (solid red), $1^-$ (dotted blue) and $2^-$ (dash-dotted green) multipoles as well as their total sum (solid black). }\label{fig:sdr}
\end{figure*}

%GT-IAR centroid difference
In Fig. \ref{fig:skin}, we show the centroid energy difference between the GTR, calculated using only the direct spin-flip transitions, and IAR for even-even tin isotopes in the range $A = 112-130$ for temperatures $T = 0,0.5,0.9,1.5$ and 2.0 MeV. The dependence of centroid energy differences on mass number is almost linear, although some deviation can be noticed for $T = 0.9$ and 1.5 MeV near the closed neutron shell. At zero temperature, our model reproduces the experimental data from Ref.~\cite{Pham1995_PRC51-526} well, with the largest difference being within 1 MeV. The centroid energy difference displays sensitivity on the isoscalar pairing strength $V^{is}$, and an overall best agreement was obtained for $V^{is} = 1.5$ (cf. Ref.~\cite{Vale2021_PRC103-064307}), which is adopted in the present calculations. Note that in Ref.~\cite{BAI-CL2014_PRC90-054335} a value of $V^{is} \approx 1.1$ was determined by similar considerations with non-relativistic functionals and a $\delta$ pairing force. From Fig. \ref{fig:skin}, it is seen that already at $T = 0.9$ MeV we get a temperature  effect  on the centroid energy difference of $\approx$ 0.5 MeV up to $A=128$, while almost no change in centroid energy difference is obtained by increasing the temperature from $T = 0.9$ up to 2.0 MeV. This temperature dependence is changed for $A = 128$ and $A = 130$ where the centroid energy differences are less affected at $T=0.9$ MeV.  We have already seen in Fig. \ref{fig:ias} that the IAS strength is almost temperature independent showing at most $ \approx 0.1$ MeV shift to lower excitation energies at $T = 1.5 $ MeV across the considered isotopic chain, and hence, only the GTR excitation energy influences the temperature dependence of the centroid energy differences. At $T=0.9$ MeV, the temperature effect plays an important role on GT strength through pairing collapse for mid-shell nuclei, however, for $^{128,130}$Sn, the pairing collapse doesn't have much influence on GT excitations, since the pairing correlations are weak for $^{128,130}$Sn that are close to the doubly magic nucleus $^{132}$Sn, and hence don't play important roles on the GT excitations, which can be seen in Fig. \ref{fig:gt} as well.

Finally, the spin-dipole (SD) excitations are obtained by setting $S=1$ and $L=1$ in the matrix element of the external field operator defined in Eq. (\ref{eq:external_matrix_element}). The matrix element can thus be coupled to $J^\pi = 0^-, 1^-$ and $2^-$. It can be shown that the SD sum rule is directly connected to the difference between neutron and proton mean-square-root radii, thus providing valuable information on neutron skin thickness~\cite{Ring1980}. The temperature evolution of SD excitation strength distributions is shown in Fig. \ref{fig:sdr} calculated with the DD-PCX interaction for ${}^{112}$Sn, ${}^{120}$Sn and ${}^{128}$Sn in the temperature range $T = 0 - 1.5$ MeV. In the figure, we plot separately contributions from $0^-, 1^-$ and $2^-$ multipolarities together with their total sum. The same strength of isoscalar pairing $V^{is}=1.5$ is used for unnatural-parity transitions $0^-$ and $1^-$ as was for GT transitions.  It is observed that the SD strength has considerably richer structure compared to the previously discussed GT and IAS.   The $2^-$ strength function shows fragmented structure while the strength of $0^-$ and $1^-$ is mostly concentrated within the main peak, due to considerably larger number of 2 q.p. excitations contributing to $2^-$ transitions, which agrees with the results from Refs.~\cite{Sagawa2007_PRC76-024301,Fracasso2007_PRC76-044307}. For all considered nuclei and temperatures, the centroid of $2^-$ transitions has the lowest energy, while the main peaks of $0^-$ and $1^-$ are $\approx 8$ MeV higher. This can be confirmed by examining Tab. \ref{tab:centroids_sdr}, where we show the $0^-, 1^-$ and $2^-$ centroid energies at $T = 0$ and $T = 1.5$ MeV together with the centroid energies of unperturbed strength. By examining Tab. \ref{tab:centroids_sdr}, it is seen that the inequality $E_{cent.}(0^-) > E_{cent}(1^-) > E_{cent}(2^-)$ is valid at $T = 0$ and $T = 1.5$ MeV for both full and unperturbed strength of all considered nuclei. A shift of $\approx$ 4 MeV of the centroid energy compared to the unperturbed one is achieved by the repulsive residual interaction. At $T = 1.5$ MeV, there is no significant shift in the centroid energy compared to the zero-temperature case, however, by examining Fig. \ref{fig:sdr}, moderate changes of the low-lying SD strength and the main peak can be noticed.

\begin{table}
\centering
\caption{The spin-dipole centroid energy $E_{cent.}$ of $0^-, 1^-$ and $2^-$ multipoles for ${}^{112}$Sn, ${}^{120}$Sn and ${}^{128}$Sn at temperatures T = 0 and T = 1.5 MeV. Values in parentheses denote centroid energies of unperturbed strength functions.  }\label{tab:centroids_sdr}
\begin{tabular}{cccc}
\hline
\hline
T = 0 MeV & $E_{cent.}(0^-)$ &  $E_{cent.}(1^-)$ & $E_{cent.}(2^-)$ \\
\hline
${}^{112}$Sn & 34.76(31.61) & 33.05(29.56) & 29.17(26.81) \\
${}^{120}$Sn & 31.54(28.22) & 29.89(25.99) & 24.75(22.10) \\
${}^{128}$Sn & 29.48(25.27) & 27.68(22.91) & 21.99(17.96) \\
\hline
T = 1.5 MeV & $E_{cent.}(0^-)$ &  $E_{cent.}(1^-)$ & $E_{cent.}(2^-)$ \\
\hline
${}^{112}$Sn & 34.75(31.07) & 32.78(28.91) & 29.16(26.28) \\
${}^{120}$Sn & 31.52(27.58) & 29.60(25.29) & 24.72(21.28) \\
${}^{128}$Sn & 29.28(24.76) & 27.34(22.40) & 21.88(18.52) \\
\hline
\end{tabular}
\end{table}

For ${}^{112}$Sn at zero temperature, the main peak is located at $E = 31.68$ MeV for  $0^-$ and  at $E = 31.33$ MeV for $1^-$. Both peaks are dominated by $(\nu 1 g_{9/2}, \pi 1h_{9/2})$ and $(\nu 1 f_{7/2}, \pi 1g_{7/2})$ transitions. With increasing temperature, the structure of these peaks in ${}^{112}$Sn remains unchanged. The $2^-$ excitation has also a peak composed of the above mentioned transitions at $E = 30.13$ MeV, however, it also has stronger peaks located lower in excitation energy with most contributions from $(\nu 1 h_{11/2}, \pi 1g_{7/2})$, $(\nu 1 g_{7/2}, \pi 1h_{11/2})$, $(\nu 2 p_{1/2}, \pi 2d_{5/2})$, and $(\nu 1 g_{9/2}, \pi 1h_{11/2})$ transitions. With increasing temperature, it can be noticed that the low-lying peak of the $2^-$ transition strength at $E \approx 9$ MeV disappears due to vanishing of pairing correlations, while the peaks at $E \approx 17$ and 34 MeV (at $T = 0$ MeV) increase in strength. However, the overall shape of the SDR, apart from a small shift of $\approx 0.5$ MeV to lower excitation energies, remains almost unchanged up to $T = 1.5$ MeV.

For ${}^{120}$Sn at zero temperature, the strength of the main peaks in the $0^-$ and $1^-$ components is fragmented due to the strong mixing of $(\nu 1 g_{9/2}, \pi 1h_{9/2})$, $(\nu 1 f_{7/2}, \pi 1g_{7/2})$, and $(\nu 1h_{11/2}, \pi 1i_{11/2})$ transitions, a similar effect as described for the GT strength in ${}^{120}$Sn. By vanishing of the pairing correlations at T $\approx$ 0.9 MeV, the fragmentation is reduced and the previously mentioned transitions start to contribute coherently to the SDR peak.  ${}^{128}$Sn exhibits a similar SD structure as previously described for ${}^{112}$Sn and ${}^{120}$Sn, although with a significantly increased strength at lower excitation energies, related to the increase in the neutron chemical potential with adding neutrons. The $0^-$ and $1^-$ peaks are now dominated by $(\nu 1 f_{7/2}, \pi 1g_{7/2})$, $(\nu 1 g_{9/2}, \pi 1h_{9/2})$, and $(\nu 1h_{11/2}, \pi 1i_{11/2})$ transitions, with an increased contribution of the $(\nu 1h_{11/2}, \pi 1i_{11/2})$ transition compared to ${}^{120}$Sn. Similar as ${}^{112}$Sn, for the $2^-$ component of ${}^{120}$Sn and ${}^{128}$Sn, as the temperature increases, the low-lying strength decreases and shifts slightly to lower excitation energies.

\section{Conclusion}\label{sec:conclusion}

In this work, we have developed the finite-temperature linear response theory based on the FT-RHB model, and applied it to the calculation of spin-isospin excitations in tin isotopes at finite-temperatures. Our approach employs the point-coupling relativistic EDFs, such as DD-PC1 and DD-PCX, for the calculation of  both the mean-field potential in the ground state and the residual $ph$ interaction in the FT-QRPA approach. The same form of the separable pairing interaction was also used both for the $pp$ interaction in FT-RHB  and for the residual $pp$ interaction in FT-QRPA. In the ground-state calculation, no proton-neutron mixing is assumed, so only the isovector ($T=1$) component of the pairing interaction contributes, while in the residual interaction both the isovector ($T=1$) and the isoscalar ($T=0$) pairing interactions can contribute.

The implementation of the linear response FT-PNQRPA was compared with the conventional matrix QRPA at zero-temperature~\cite{Vale2021_PRC103-064307} and also at finite-temperature (based on the FT-HBCS ground-state)~\cite{Yuksel2020_PRC101-044305}, which successfully reproduces the results of both matrix implementations. The linear response QRPA based on separable forces avoids the diagonalization of large QRPA matrices (especially at finite-temperature), so it provides a fast and efficient method for obtaining the spin-isospin excitation strength functions.

We have shown that the temperature has almost no effects on IAS excitations, and the whole IAS strength remains concentrated in one single peak at finite-temperature.  For GT transitions, temperature effects are particularly important for ${}^{120}$Sn and ${}^{124}$Sn, where with vanishing pairing correlations, the fragmentation of the main peak disappears at $T \approx 0.9$ MeV. We also studied the temperature dependence of the GTR and IAR centroid difference, demonstrating a visible effect already at T = 0.9 MeV. At zero temperature by setting $V^{is} = 1.5$, an agreement with experimental data from Ref.~\cite{Pham1995_PRC51-526} was obtained within 1 MeV. Lastly, we have studied the temperature evolution of the SD excitation strengths in ${}^{112}$Sn, ${}^{120}$Sn, and ${}^{128}$Sn. A clear hierarchy of $0^-, 1^-$ and $2^-$ transitions was confirmed also at finite-temperatures, where the centroid of $0^-$ excitations is located at highest and the centroid of $2^-$ excitations at lowest excitation energy. Moderate effects on the shape of the SD excitations were found at finite-temperature, mainly related to the reduction of strength in low-lying peaks with vanishing pairing correlations, and a removal of the SDR fragmentation in ${}^{120}$Sn, as was also confirmed for GT strength at $T \approx 0.9$ MeV.

As was emphasized in the introduction, the study of the spin-isospin response is important for the calculation of weak-interaction rates which serve as inputs in many astrophysical scenarios (e.g. r-process and core-collapse supernovae). The efficiency in calculating the excitation strengths within the linear response formalism makes large-scale calculations of electron capture, $\beta-$decay and neutrino-nucleus reactions at zero and finite-temperature feasible. Besides, the linear response implementation of the QRPA has its advantage in saving computational effort when extending to the description of deformed nuclei, where the angular momentum $J$ is no longer a good quantum number, thus drastically increasing the space of available 2 q.p. excitations. We leave the implementation of axially-deformed QRPA in the linear-response formalism for the future.

\section{Acknowledgements}
This work is supported by the QuantiXLie Centre of Excellence, a project co financed by the Croatian Government and European Union through the European Regional Development Fund, the Competitiveness and Cohesion Operational Programme (KK.01.1.1.01.0004). This article is based upon work from the ChETEC COST Action (CA16117), supported by COST (European Cooperation in Science and Technology).
Y. F. N. acknowledges the support from National Natural Science Foundation of China under Grant No. 12075104, and the Fundamental Research Funds for the Central Universities under Grant No. Lzujbky-2019-11. P. R. acknowledges support from the Deutsche Forschungsgemeinschaft (DFG, German Research Foundation) under Germany's Excellence Strategy EXC-2094-390783311, ORIGINS.
\appendix

\section{Separable channels of the DD-PC1 and DD-PCX interactions}\label{sec:app_a}

Since we are considering the charge-exchange excitations, the only terms in the residual interaction of the point-coupling functionals that can contribute (due to charge-conservation) are
\begin{itemize}

\item the isovector-vector (TV) interaction
\begin{equation}\label{eq:res1}
\langle a b |V_{TV} | c d \rangle = -  \alpha_{TV}(\rho_v) (\bar{\psi}_a \boldsymbol{\tau} \gamma_\mu \psi_c) \cdot (\bar{\psi}_b \boldsymbol{\tau} \gamma^\mu \psi_d) \delta(\boldsymbol{r}_1 - \boldsymbol{r}_2),
\end{equation}

\item the isovector-pseudovector (TPV) interaction
\begin{equation}\label{eq:res2}
\langle a b |V_{TPV} | c d \rangle= g_0 ( \bar{\psi}_a\gamma_0 \gamma_5 \gamma_\mu \boldsymbol{\tau} \psi_c) \cdot ( \bar{\psi}_b \gamma_0 \gamma_5 \gamma^\mu \boldsymbol{\tau} \psi_d) \delta(\boldsymbol{r}_1 - \boldsymbol{r}_2),
\end{equation}

\end{itemize}
where $\cdot$ implies integration over $\boldsymbol{r}_1, \boldsymbol{r}_2$ as well as summation over $\mu$. The parameter $g_0$ is the TPV interaction coupling constant. We note that no TPV term is present in the Lagrangian density of the point-coupling functionals in the ground-state, therefore $g_0$ should be determined from the excited state properties. Its strength is $g_0 = 0.734$ for the DD-PC1 interaction and $g_0 = 0.621$ for DD-PCX as determined by reproducing the experimental GT${}^-$ centroid energy in ${}^{208}$Pb \cite{deni_paper}. The Dirac spinors in the central field with spherical symmetry have the form~\cite{Gambhir1990_APNY198-132}
\begin{equation}\label{eq:dirac_spinor}
\langle \boldsymbol{r}| \psi \rangle = \begin{pmatrix}
f_i(r) \left[\chi_{1/2} \otimes Y_l(\Omega) \right]_{j m } \\
i g_i(r) \left[\chi_{1/2} \otimes Y_{\tilde{l}}(\Omega) \right]_{j m } \\
\end{pmatrix},
\end{equation}
where $f_i(r)$ ($g_i(r)$) are upper (lower) components of the Dirac spinor, $j$ labels the total angular momentum with projection $m$ and orbital angular momentum $l$ ($\tilde{l}$) for upper (lower) components, while $\chi_{1/2 m_s}$ are spin $1/2$ wavefunctions with projection $m_s$. The Dirac spinors are expanded in $N_{osc}$ ($\tilde{N}_{osc}$) harmonic oscillator shells for upper(lower) components~\cite{Gambhir1990_APNY198-132}
\begin{align}
f_i(r) = \sum \limits_{n = 0}^{N_{osc}} f_n^{(i)} R_{n l_i} (r,b), \quad g_i(r) = \sum \limits_{n = 0}^{\tilde{N}_{osc}} g_{n }^{(i)} R_{n \tilde{l}_i}(r,b),
\end{align}
where $R_{n l_i} (r,b)$ are radial harmonic oscillator wavefunctions, and $b$ the oscillator length defined as $b = \sqrt{\hbar/m \omega_0}$, where $m$ is the bare nucleon mass and $\hbar \omega_0$ the oscillator frequency~\cite{Gambhir1990_APNY198-132}. The residual interaction $V_{ph}$ can be written as a product of separable terms
\begin{equation}\label{eq-A4}
\langle k_1 k_2 | V_{ph} | k_3 k_4 \rangle =\int r^2 dr \int r^{\prime 2} dr^\prime \sum \limits_c Q_{c k_1 k_3}(r) v_{c}(r, r^\prime) Q_{c k_2 k_4}(r^\prime),
\end{equation}
where $c$ is the interaction channel index and $v_c(r,r^\prime)$ contains the radial dependence. For point coupling models we have $v_{c}(r, r^\prime)\sim\delta(r-r^\prime)$ and the remaining radial integral can be represented as a sum over the meshpoints $r_i$. In this case we can combine $i$ with the channel index $c$ to $\rho=(i,c)$ and obtain the matrix elements  (\ref{eq-A4}) in the separable form of Eq. (\ref{eq:full_ham}) with appropriate coupling constants $\chi_\rho$. $k_1,k_2,k_3,k_4$ denote single-particle states in the basis of a spherical harmonic oscillator. Separable channels can be distinguished between natural parity and unnatural parity transitions. Introducing $f_{n l_i}(r) \equiv f_n^{(i)}R_{n l_i}(r,b)$, they are given by
\begin{widetext}
\begin{itemize}

\item natural parity transitions
\begin{equation}
Q_{1 k_1 k_3}(r)= f_{n_{k_1} l_{k_1}}(r) f_{n_{k_3} l_{k_3}}(r)  \langle l_{k_1} j_{k_1} || Y_{J} \left(\Omega\right) \boldsymbol{\tau}|| l_{k_3} j_{k_3} \rangle   + g_{n_{k_1} \tilde{l}_{k_1}}(r) g_{n_{k_3} \tilde{l}_{k_3}}(r) \langle \tilde{l}_{k_1} j_{k_1} || Y_{J} \left(\Omega\right) \boldsymbol{\tau}|| \tilde{l}_{k_3} j_{k_3} \rangle ,
\end{equation}
\begin{equation}
Q_{2 k_1 k_3}(r) =  f_{n_{k_1} l_{k_1}}(r) g_{n_{k_3} \tilde{l}_{k_3}} (r) \langle l_{k_1} j_{k_1} || \left[ \sigma_S Y_{J-1} \left(\Omega\right) \right]_J \boldsymbol{\tau} || \tilde{l}_{k_3} j_{k_3} \rangle  -  g_{n_{k_1} \tilde{l}_{k_1}}(r) f_{n_{k_3} l_{k_3}} (r) \langle \tilde{l}_{k_1} j_{k_1} || \left[ \sigma_S Y_{J-1} \left(\Omega\right) \right]_J \boldsymbol{\tau} || l_{k_3} j_{k_3} \rangle,
\end{equation}
\begin{equation}
Q_{3 k_1 k_3}(r) =  f_{n_{k_1} l_{k_1}}(r) g_{n_{k_3} \tilde{l}_{k_3}} (r) \langle l_{k_1} j_{k_1} || \left[ \sigma_S Y_{J+1} \left(\Omega\right) \right]_J \boldsymbol{\tau} || \tilde{l}_{k_3} j_{k_3} \rangle  -  g_{n_{k_1} \tilde{l}_{k_1}}(r) f_{n_{k_3} l_{k_3}} (r) \langle \tilde{l}_{k_1} j_{k_1} || \left[ \sigma_S Y_{J+1} \left(\Omega\right) \right]_J \boldsymbol{\tau} || l_{k_3} j_{k_3} \rangle,
\end{equation}
\begin{equation}
Q_{4 k_1 k_3}(r)= f_{n_{k_1} l_{k_1}}(r) f_{n_{k_3} l_{k_3}}(r) \langle l_{k_1} j_{k_1} || \left[ \sigma_S Y_{J} \left(\Omega\right) \right]_J \boldsymbol{\tau}|| l_{k_3} j_{k_3} \rangle   + g_{n_{k_1} \tilde{l}_{k_1}}(r) g_{n_{k_3} \tilde{l}_{k_3}}(r) \langle \tilde{l}_{k_1} j_{k_1} || Y_{J} \left(\Omega\right) \boldsymbol{\tau}|| \tilde{l}_{k_3} j_{k_3} \rangle,
\end{equation}
with $v_1(r,r^\prime) = \frac{\alpha_{TV}(r)}{r^2}\delta(r-r^\prime), v_2(r,r^\prime) = -\frac{\alpha_{TV}(r)}{r^2}\delta(r-r^\prime), v_3(r,r^\prime) = -\frac{\alpha_{TV}(r)}{r^2}\delta(r-r^\prime), v_4(r,r^\prime) = -\frac{g_0}{r^2}\delta(r-r^\prime)$.

\item unnatural parity
\begin{equation}
Q_{1 k_1 k_3}(r) =   g_{n_{k_1} \tilde{l}_{k_1}}(r) f_{n_{k_3} l_{k_3}} (r) \langle \tilde{l}_{k_1} j_{k_1} || \left[ \sigma_S Y_{J} \left(\Omega\right) \right]_J \boldsymbol{\tau} || l_{k_3} j_{k_3} \rangle - f_{n_{k_1 } l_{k_1}}(r) g_{n_{k_3} \tilde{l}_{k_3}} (r) \langle l_{k_1} j_{k_1} || \left[ \sigma_S Y_{J} \left(\Omega\right) \right]_J \boldsymbol{\tau} || \tilde{l}_{k_3} j_{k_3} \rangle,
\end{equation}
\begin{equation}
Q_{2 k_1 k_3}(r) =  f_{n_{k_1} l_{k_1}}(r) g_{n_{k_3} \tilde{l}_{k_3}} (r) \langle l_{k_1} j_{k_1} || Y_J \left(\Omega\right) \boldsymbol{\tau} || \tilde{l}_{k_3} j_{k_3} \rangle -  g_{n_{k_1} \tilde{l}_{k_1}}(r) f_{n_{k_3} l_{k_3}} (r) \langle \tilde{l}_{k_1} j_{k_1} ||  Y_{J} \left(\Omega\right) \boldsymbol{\tau} || l_{k_3} j_{k_3} \rangle ,
\end{equation}
\begin{equation}
Q_{3 k_1 k_3}(r)= f_{n_{k_1} l_{k_1}}(r) f_{n_{k_3} l_{k_3}}(r) \langle l_{k_1} j_{k_1} || \left[ \sigma_S Y_{J-1} \left(\Omega\right) \right]_J \boldsymbol{\tau}|| l_{k_3} j_{k_3} \rangle   + g_{n_{k_1} \tilde{l}_{k_1}}(r) g_{n_{k_3} \tilde{l}_{k_3}}(r) \langle \tilde{l}_{k_1} j_{k_1} || Y_{J-1} \left(\Omega_{1}\right) \boldsymbol{\tau}|| \tilde{l}_{k_3} j_{k_3} \rangle ,
\end{equation}
\begin{equation}
Q_{4 k_1 k_3}(r)= f_{n_{k_1} l_{k_1}}(r) f_{n_{k_3} l_{k_3}}(r) \langle l_{k_1} j_{k_1} || \left[ \sigma_S Y_{J-1} \left(\Omega\right) \right]_J \boldsymbol{\tau}|| l_{k_3} j_{k_3} \rangle   + g_{n_{k_1} \tilde{l}_{k_1}}(r) g_{n_{k_3} \tilde{l}_{k_3}}(r) \langle \tilde{l}_{k_1} j_{k_1} || Y_{J-1} \left(\Omega\right) \boldsymbol{\tau}|| \tilde{l}_{k_3} j_{k_3} \rangle,
\end{equation}
with $v_1(r,r^\prime) = -\frac{g_0}{r^2}\delta(r-r^\prime), v_2(r,r^\prime) = \frac{g_0}{r^2}\delta(r-r^\prime), v_3(r,r^\prime) = -\frac{\alpha_{TV}(r)}{r^2}\delta(r-r^\prime), v_4(r,r^\prime) = -\frac{g_0}{r^2}\delta(r-r^\prime)$.

\end{itemize}

\end{widetext}
The spin rank is either $S=0$ or $S=1$ therefore the rank of spherical harmonics is $J, J \pm 1$ so that the total matrix element can be coupled to $J$. The isospin Pauli matrix is denoted by $\boldsymbol{\tau}$, while the spin matrix is $\sigma_S$. There are only 4 channels for both cases of natural parity. Finally, $ph$ separable matrix elements are transformed from the harmonic oscillator basis to the Gauss-Hermite coordinate mesh.

Two-body matrix elements of residual pairing interaction in the basis of spherical harmonic oscillator are calculated as
\begin{equation}\label{eq:mat_el_sep_pair}
\langle n_{k_1} l_{k_1} j_{k_1}, n_{k_2} l_{k_2} j_{k_2} |\hat{V}_{pp} (\boldsymbol{r}_1, \boldsymbol{r}_2, \boldsymbol{r}_1^\prime, \boldsymbol{r}_2^\prime) | n_{k_3} l_{k_3} j_{k_3}, n_{k_4} l_{k_4} j_{k_4} \rangle,
\end{equation}
where we assume the separable interaction in Eq. (\ref{eq-15}):
\begin{equation}
\hat{V}_{pp}(\boldsymbol{r}_1, \boldsymbol{r}_2, \boldsymbol{r}_1^\prime, \boldsymbol{r}_2^\prime) = -G \delta(\boldsymbol{R}-\boldsymbol{R}^\prime) P(r) P(r^\prime) \frac{1}{2} (1- P^r P^\sigma P^\tau ),
\end{equation}
with $P^r, P^\sigma, P^\tau$ being coordinate, spin and isospin exchange operators respectively, other expressions being defined in Sec. \ref{sec:theoretical_formalism}. By calculating matrix element in Eq. (\ref{eq:mat_el_sep_pair}) and coupling to good angular momentum $J$ residual pairing matrix elements assume the separable form
\begin{equation}\label{eq:separable_pairing}
V^J_{k_1 k_2,k_3 k_4} = -G f  \sum \limits_{N L S} V^{N L S J}_{k_1 k_2} V^{N L S J}_{k_3 k_4},
\end{equation}
where we define the separable terms as
\begin{align}
\begin{split}
V^{N L S J}_{k_1 k_2} &= \frac{\hat{L} \hat{S}}{2^{3/2} \pi^{3/4} b^{3/2}} \frac{(1-\alpha^2)^n}{(1+\alpha^2)^{n+3/2}} \times \\ &\times \frac{(2n+1)!}{2^n n!} \hat{j}_{k_1} \hat{j}_{k_2} \begin{Bmatrix}
l_{k_2} & 1/2 & j_{k_2} \\
l_{k_1} & 1/2 & j_{k_1} \\
L & S & J \\
\end{Bmatrix} M^{N L n 0}_{n_{k_1} l_{k_1} n_{k_2} l_{k_2}},
\end{split}
\end{align}
with $\alpha = a/b$, $a$ being the width parameter in Eq. (\ref{eq-16}) and $b$ the harmonic oscillator constant. $M^{N L n 0}_{n_{k_1} l_{k_1} n_{k_2} l_{k_2}}$ are the Talmi-Moschinsky brackets~\cite{TIAN-Y2009_PRC79-064301}. We use the usual abbreviation $\hat{j} = \sqrt{2j +1}$. Due to constraints imposed by coupling charge-exchange channel and exchange operators of Eq. (\ref{eq:mat_el_sep_pair})
\begin{equation}
f = \begin{cases}
    1,& \text{if } T = 1, S = 0\\
    V^{is},& \text{if } T = 0, S = 1 \\
    0 & \text{otherwise}
\end{cases}.
\end{equation}
For the charge-exchange channel total isospin operator can assume values $T = 0,1$. In Eq. (\ref{eq:separable_pairing}) we restrict the summation up to maximum $N = 8$. For DD-PCX interaction where $G_p \neq G_n$ [cf. Sec. \ref{sec:theoretical_formalism}] we use their average $G = (G_p + G_n)/2$.

\section{Separable channels in proton-neutron quasiparticle basis}\label{sec:appb}

The single-particle operator $D_\rho$ defined in Eq. (\ref{eq:full_ham}) in the proton-neutron basis is given by
\begin{equation}\label{eq:original_op}
D^\dag =  \sum_{pn} Q^*_{pn} c_p^\dag c_n,
\end{equation}
where for simplicity we drop the $\rho$ channel label. Separable matrix elements of point-coupling interaction are denoted as $Q_{pn}$ while $c_{p(n)}$ and $c^\dag_{p(n)}$ are proton (neutron) annihilation and creation operators respectively. Note that in the proton-neutron basis the number of separable channels is doubled to account for second term in Eq. (\ref{eq:W_matrix}). The Bogoliubov transformation between fermion operators to the quasiparticle basis assuming spherical symmetry is defined by~\cite{Suhonen2007}
\begin{align}
c_{k j -m} &= \sum \limits_{l} U_{k l}^j \beta_{l j -m} + (-)^{j+m}V^{j *}_{k l} \beta^\dag_{l j m}, \\
c^\dag_{k j m} &= \sum \limits_{l} (-)^{j-m} V_{k l}^j \beta_{l j -m} + U^{j *}_{k l} \beta^\dag_{l j m},
\end{align}
where indices $k,l$ denote single-particle states in harmonic oscillator basis while $j$ is the total angular momentum of the state and $m$ its projection. In the above, $\beta_{l j m}, \beta^\dag_{l j m}$ denote annihilation and creation quasiparticle operators. We notice that matrices $U$ and $V$ are independent of projection $m$. In order to couple the quasiparticle operators to good $J$ and projection $M$ we define the couplings~\cite{Ring1980,Suhonen2007}
\begin{align}
&[\beta_{p j}^\dag \beta_{n j^\prime}^\dag]_{J M} = \sum \limits_{m m^\prime} C^{J M}_{j m j^\prime m^\prime} \beta^\dag_{p j m} \beta_{n j^\prime m^\prime}^\dag, \label{eq:fermion_couplings_1} \\
& [\tilde{\beta}_{p j} \tilde{\beta}_{n j^\prime}]_{J M} = -(-)^{J +M}\sum \limits_{m m^\prime} C^{J -M}_{j m j^\prime m^\prime} \beta_{p j m} \beta_{n j^\prime m^\prime}, \\
&[\beta^\dag_{p j } \otimes \tilde{\beta}_{n j^\prime }]_{J M} = \sum \limits_{m m^\prime} (-)^{j^\prime - m^\prime} C^{J M}_{j m j^\prime -m^\prime} \beta^\dag_{p j m} \beta_{n j^\prime m^\prime}, \\
& [\tilde{\beta}_{p j } \otimes \beta^\dag_{n j^\prime }]_{J M} = -(-)^{J+M}\sum \limits_{m m^\prime} (-)^{j^\prime - m^\prime} C^{J -M}_{j m j^\prime -m^\prime} \beta_{p j m} \beta^\dag_{n j^\prime m^\prime},\label{eq:fermion_couplings_2}
\end{align}
where $\tilde{\beta}_{jm} = (-)^{j+m}\beta_{j-m}$. Single-particle operator in Eq.  (\ref{eq:original_op}) can therefore be transformed to the spherical q.p. basis as
\begin{align}
\begin{split}
\hat{Q}_{pn} &= \sum \limits_{pn; j j^\prime; m m^\prime} Q_{p j m; n j^\prime m^\prime} c^\dag_{p j m} c_{n j^\prime m ^\prime} \\
&=  \sum \limits_{pn; j j^\prime; m m^\prime} Q_{p j m; n j^\prime m^\prime} \left( (-)^{j-m} V_{p \pi}^j U_{n \nu}^{j^\prime} \beta_{\pi j -m} \beta_{\nu j^\prime m^\prime} \right. \\
&+ (-)^{j + j^\prime - m -m^\prime} V_{p \pi}^j V^{j^\prime *}_{n \nu} \beta_{\pi j -m} \beta^\dag_{\nu j^\prime -m^\prime}  \\
&  + U^{j *}_{p \pi} U_{n \nu}^{j^\prime} \beta^\dag_{\pi j m} \beta_{\nu j^\prime m^\prime}  \\
&+ \left. (-)^{j^\prime - m^\prime} U^{j *}_{p \pi} V^{j^\prime *}_{n \nu} \beta^\dag_{\pi j m} \beta^\dag_{\nu j^\prime -m^\prime} \right).
\end{split}
\end{align}
By performing the coupling of the $ph$ matrix element defined as
\begin{equation}
Q_{p j m; n j^\prime m^\prime} = \sum \limits_{JM} Q^J_{p j; n j^\prime} (-)^{j^\prime - m^\prime} C^{J M}_{j m j^\prime -m^\prime},
\end{equation}
and inserting in above derivation with use of Eqs. (\ref{eq:fermion_couplings_1} - \ref{eq:fermion_couplings_2}) we get
\begin{align}
\begin{split}
\hat{Q}_{pn}
&= \sum \limits_{p \pi n\nu j j^\prime J M}  Q^J_{p j; n j^\prime}  \left(   V_{p \pi}^j U_{n \nu}^{j^\prime} [ \tilde{\beta}_{\pi j }  \tilde{\beta}_{\nu j^\prime}]_{J M}   \right. \\
&-   V_{p \pi}^j V^{j^\prime *}_{n \nu} [\tilde{\beta}_{\pi j } \otimes \beta^\dag_{\nu j^\prime }]_{J M}   \\
& \left. +  U^{j *}_{p \pi} U_{n \nu}^{j^\prime} [\beta^\dag_{\pi j } \otimes \tilde{\beta}_{\nu j^\prime }]_{J M} +   U^{j *}_{p \pi} V^{j^\prime *}_{n \nu} [\beta^\dag_{\pi j } \beta^\dag_{\nu j^\prime}]_{J M} \right). \\
\end{split}
\end{align}
In the matrix notation introduced in Sec. \ref{sec:theoretical_formalism}
\begin{equation}
\hat{Q}_{pn} = \sum \limits_{\pi \nu} \begin{pmatrix}
U^\dag Q^J U & U^\dag Q^J V^* \\
V^T Q^J U & -V^T Q^J V^*
\end{pmatrix}_{\pi \nu} [a_\pi^\dag a_\nu]_{J M}.
\end{equation}
For the residual pairing interaction, the single-particle operator $D_\rho$ assumes the following form
\begin{equation}
D^\dag = \sum \limits_{pn} V^{N L S J}_{p n} c_p^\dag c_n^\dag ,
\end{equation}
which when rewritten in q.p. basis and coupled to total angular momentum $J$ yields Eq. (\ref{eq:pp_pn}).

\section{Calculating (quasi)particle-(quasi)hole contributions within the linear response theory}\label{sec:appc}

In order to gain additional information from the response function we need to establish correspondence between the discrete matrix FT-QRPA modes and linear response FT-QRPA. Matrix FT-QRPA equations can be obtained by rewriting Eq. (\ref{eq:lin_res_matrix_xy}) in the form
\begin{align}
\begin{split}
&\delta \mathcal{R}_{\pi \nu} = \frac{f_{\nu} - f_\pi}{\omega - E_\pi + E_{\nu}} \biggl\{ F_{\pi \nu}  \biggr.\\
&+ \biggl. \sum \limits_c \int r^2 dr v_c(r) Q_{c \pi \nu}(r) \int r^{\prime 2} dr^\prime f(r^\prime) R_{c F}(\omega; r^\prime) \biggr\},
\end{split}
\end{align}
where we have used the definition of the two-particle matrix element [cf. Eq. (\ref{eq:effective_interacton_W})]
\begin{equation}
\mathbb{W}_{\pi \nu \pi^\prime \nu^\prime} = \int r^2 dr Q_{c \pi \nu}(r) v_c(r) Q_{c \pi^\prime \nu^\prime}^*(r),
\end{equation}
with separable interaction matrix element $Q_{c \pi \nu}$ for the channel $c$ of the residual interaction, while the radial dependence is written as $v_c(r)$. The response function $R_{cF}(\omega;r)$ can be calculated from the Bethe-Salpeter equation
\begin{align}
\begin{split}
&R_{c^\prime F}(\omega;r^\prime) = R^0_{c^\prime F}(r^\prime) \\
&+ \sum \limits_{c c^{\prime \prime}} \int r^2 dr r^{\prime \prime 2} dr^{\prime \prime} R^0_{c^\prime c}(r^\prime, r) v_{c, c^{\prime \prime}}(r, r^{\prime \prime})R_{c^{\prime \prime} F}(\omega; r^{\prime \prime}),
\end{split}
\end{align}
where $R^0_{c^\prime c}(r^\prime, r)$ is obtained from Eq. (\ref{eq:reduced_unp_resp}), while
\begin{equation}
R^0_{c^\prime F}(r^\prime) =  \sum \limits_{\pi \nu} \frac{f_{\nu}-f_\pi}{\omega - E_\pi + E_{\nu} + i \eta} Q^*_{c^\prime \pi \nu}(r^\prime) F_{\pi \nu}(r^\prime).
\end{equation}
We can now define the linear response amplitudes $\delta \mathcal{R}_{\pi \bar{\nu}} = X_{\pi \nu}(\omega), \delta \mathcal{R}_{\bar{\pi} \nu} = Y_{\pi \nu}(\omega), \delta \mathcal{R}_{\pi \nu} = P_{\pi \nu}(\omega)$ and $\delta \mathcal{R}_{\bar{\pi} \bar{\nu}} = Q_{\pi \nu}(\omega)$. Using the formalism developed in Ref.~\cite{Sommermann1983_APNY151-163} one can connect finite-temperature linear response function $\mathbb{R}$ with the eigenvectors of the matrix FT-QRPA approach. Employing the normalization of the FT-QRPA eigenvectors and contour integration around suitably chosen loop $C_i$ the linear response amplitudes can be used to calculate FT-QRPA eigenvectors of the $i-$th mode
\begin{align}
X^i_{\pi \nu} &= e^{-i \theta} |\langle i | \hat{F} | 0 \rangle|^{-1} \frac{1}{2 \pi i} \oint_{C_i} X_{\pi \nu} (\omega) d \omega, \label{eq:eigen1} \\
Y^i_{\pi \nu} &= e^{-i \theta} |\langle i | \hat{F} | 0 \rangle|^{-1} \frac{1}{2 \pi i} \oint_{C_i} Y_{\pi \nu} (\omega) d \omega, \\
P^i_{\pi \nu} &= e^{-i \theta} |\langle i | \hat{F} | 0 \rangle|^{-1} \frac{1}{2 \pi i} \oint_{C_i} P_{\pi \nu} (\omega) d \omega, \\
Q^i_{\pi \nu} &= e^{-i \theta} |\langle i | \hat{F} | 0 \rangle|^{-1} \frac{1}{2 \pi i} \oint_{C_i} Q_{\pi \nu} (\omega) d \omega, \label{eq:eigen4}
\end{align}
where we have extended the formalism of Ref.~\cite{Hinohara2013_PRC87-064309} to finite-temperature. The overall phase $e^{-i \theta}$ remains undetermined. The FT-QRPA matrix elements of the external field operator $|\langle i | \hat{F} | 0 \rangle|$ can be calculated using Eq. (\ref{eq:contour_strength}). Above system of equations can be easily discretized on the circular loop of small radius $\eta$ that encloses the $i-$th pole and integrated using Simpson's or Trapezoidal rule. Once the eigenvectors from Eqs.(\ref{eq:eigen1}-\ref{eq:eigen4}) are calculated for the particular 2 q.p. excitation its strength matrix element is obtained as in the usual matrix FT-QRPA calculations [cf. Eq. (\ref{eq:ftqrpa_excitation_me})].

\bibliographystyle{apsrev4-1}
\bibliography{bib-peter}

\end{document}